\documentclass[aps,pre,twocolumn,groupedaddress,amsmath,showpacs,eqsecnum]{revtex4}
   \usepackage{bm}
\usepackage{graphicx,amssymb}
\usepackage{bm}
\usepackage{amsmath}
\newcommand{\s}{{\text{s}}}
\newcommand{\visc}{\eta}
\newcommand{\beq}{\begin{equation}}
\newcommand{\eeq}{\end{equation}}
\newcommand{\beqa}{\begin{eqnarray}}
\newcommand{\eeqa}{\end{eqnarray}}
\newcommand{\dd}{{d}}

\newcommand{\nn}{\nonumber\\}

\begin{document}

\title{Uniform shear flow in dissipative
gases. Computer simulations of inelastic hard spheres and
(frictional) elastic hard spheres}
\author{Antonio Astillero}
\email{aavivas@unex.es}
\homepage{http://www.unex.es/eweb/fisteor/antonio/}
\affiliation{Departamento de Inform\'atica, Centro Universitario de
M\'erida, Universidad de Extremadura, E--06800 M\'erida, Badajoz,
Spain}
\author{Andr\'es Santos}
\email{andres@unex.es}
\homepage{http://www.unex.es/eweb/fisteor/andres/}
\affiliation{Departamento de F\'{\i}sica, Universidad de
Extremadura, E--06071 Badajoz, Spain}
\date{\today}
\begin{abstract}
In the preceding paper (cond-mat/0405252), we have conjectured that the main transport
properties of a dilute gas of inelastic hard spheres (IHS) can be
satisfactorily
captured by an equivalent gas of elastic hard spheres (EHS),
provided that the latter are under the action of an effective drag force
and their collision rate is reduced by a factor $(1+\alpha)/2$ (where $\alpha$ is the constant coefficient of normal restitution).
In this paper we test the above expectation in a paradigmatic nonequilibrium state, namely the simple or uniform shear flow, by
performing Monte Carlo computer simulations of the Boltzmann
equation for both classes of dissipative gases with  a dissipation range $0.5\leq
\alpha\leq 0.95$ and two values of the imposed shear rate $a$. It is observed that the evolution toward the steady state proceeds in two
stages: a
short kinetic stage (strongly dependent on the initial preparation of
the system) followed
by a  slower hydrodynamic regime that becomes increasingly less
dependent on the initial state. Once conveniently scaled, the intrinsic quantities in the hydrodynamic regime depend on time, at a
given value of $\alpha$, only through the
 reduced
shear rate $a^*(t)\propto a/\sqrt{T(t)}$, until a steady state, independent of the imposed shear rate
and of the initial preparation,  is reached.
The distortion of the steady-state velocity distribution from the local
equilibrium state is measured by the shear stress, the normal
stress differences, the cooling rate, the fourth and sixth cumulants, and the shape of the distribution itself.
In particular, the simulation results
seem to be consistent with an exponential overpopulation of the high-velocity tail.
These properties are common to both the IHS and EHS systems. In
addition, the EHS results are in general hardly distinguishable from the IHS
ones if $\alpha\gtrsim 0.7$, so that the distinct signature of the IHS gas
(higher anisotropy and overpopulation) only manifests itself at relatively high
dissipations.

\end{abstract}

\maketitle
\section{Introduction\label{sec1}}
A granular gas in the rapid flow regime is usually modeled as a
system of smooth inelastic hard spheres with a constant coefficient
of normal restitution $\alpha$. The key ingredient of this minimal model is
that  energy is not conserved in collisions: in every binary
collision, an amount of energy proportional to $1-\alpha^2$ is
transferred to the internal degrees of freedom and thus it is lost
as translational kinetic energy. Therefore, the gas ``cools''
down and the granular temperature monotonically decreases in time
unless energy is externally injected into the system to compensate
for the collisional loss. If this energy injection takes place
through the boundaries (e.g., vibrating walls, thermal walls,
sliding walls, \ldots) a nonequilibrium steady state can be reached
characterized by strong spatial gradients in basic average
quantities, such as density, mean velocity, or temperature. Kinetic theory tools can be straightforwardly extended
to granular gases and, in particular, the Boltzmann and Enskog
kinetic equations have been formulated for inelastic collisions
\cite{GS95,BDS97,vNE01}.

In the preceding paper \cite{SA04}, we have suggested that the
nonequilibrium transport properties of a genuine gas of  inelastic
hard spheres (IHS) can   be accounted for, to some extent, by an
``equivalent'' gas of elastic hard spheres (EHS). This requires the
inclusion of two basic points. First, the EHS are assumed to feel
the action of a drag force with a friction coefficient that  mimics
the collisional cooling rate
of the true IHS gas. This guarantees that the local energy balance
is approximately the same in both systems. Second, the collision rate of the
EHS gas must be decreased by a factor $\beta(\alpha)$ with respect
to that of the IHS gas at the same (local) density and temperature.
This can be interpreted under the assumption that, while the EHS have
the same mass $m$ as the IHS, they have a  diameter
$\sigma'=\beta^{1/2}\sigma$ (where henceforth we are restricting ourselves to
three-dimensional systems) smaller than the diameter $\sigma$ of the
IHS. Comparison between some basic collision integrals for IHS and
 EHS suggests the simple choice
$\beta=\frac{1}{2}(1+\alpha)$ \cite{SA04} to optimize the agreement between both descriptions.

The aim of this paper is two-fold. First, we want to assess to what extent the EHS gas
succeeds in mimicking the transport properties and other quantities of the IHS gas by
choosing the most widely studied nonequilibrium state,
namely the simple or uniform shear flow (USF). Since the USF state
is intrinsically non-Newtonian \cite{G03,SGD03}, it provides an
interesting test case to check whether or not the IHS and EHS gases
behave similarly in extreme situations far from equilibrium.
As we will see, it turns out that most of the nonequilibrium properties of both types of system are practically indistinguishable
for degrees of dissipation as large as $\alpha\approx 0.7$. For higher
dissipations, the agreement is still at least semi-quantitative. The second
goal of the paper is to carry out a rather extensive study of the most relevant properties of the USF in dissipative gases.
This study includes the unsteady or transient regime (usually neglected in
previous studies), which can be decomposed into a \textit{kinetic} stage
followed by a \textit{hydrodynamic} stage. In the steady state we
pay special attention not only to the rheological properties, but
also to the velocity cumulants and to the distribution function
itself.

The paper is organized as follows. The Boltzmann equation for both
types of system, IHS and EHS, is briefly recalled in Sec.\
\ref{sec2}, where also part of the notation is introduced. This is
followed in Sec.\ \ref{sec3} by a description of the USF and of the main quantities
of interest. Section \ref{sec4} is devoted to
some details of the numerical simulation method employed to solve
the Boltzmann equation for each class of systems. The most important part of the paper is
contained in Sec.\ \ref{sec5}, where the results for the transient and steady-state problems are presented and
discussed. When possible, the simulation data are also compared with
predictions from a simple kinetic model \cite{BDS99,SA04}. The paper
is closed by some concluding remarks in Sec.\ \ref{sec6}.

\section{Boltzmann equation\label{sec2}} In a kinetic theory
description the relevant quantity is the one-particle velocity
distribution function $f({\bf r},{\bf v};t)$. Its first five moments
define the number density
\begin{equation}
n(\mathbf{r},t)=\int \dd\mathbf{v}\, f({\bf r},{\bf v};t),
\label{b1}
\end{equation}
the nonequilibrium flow velocity
\begin{equation}
\mathbf{u}(\mathbf{r},t)=\frac{1}{n(\mathbf{r},t)}\int
\dd\mathbf{v}\, \mathbf{v} f({\bf r},{\bf v};t),
\label{b2}
\end{equation}
and the \textit{granular} temperature
\begin{equation}
T(\mathbf{r},t)=\frac{m}{3n(\mathbf{r},t)}\int \dd\mathbf{v}\,
V^2(\mathbf{r},t)f({\bf r},{\bf v};t) ,
\label{b3}
\end{equation}
where $m$ is the mass of a particle and ${\bf V}({\bf
r},t)\equiv{\bf v}-{\bf u}({\bf r},t)$ is the peculiar velocity. The hydrostatic pressure  $p=nT$  is
one third the trace of the pressure tensor $\mathsf{P}$ defined as
\begin{equation}
\mathsf{P}(\mathbf{r},t)={m}\int \dd\mathbf{v}\,
\mathbf{V}(\mathbf{r},t)\mathbf{V}(\mathbf{r},t)f({\bf r},{\bf
v};t).
\label{c1}
\end{equation}

\subsection{Inelastic hard spheres}
 The evolution of $f$ in the
low-density limit is governed by the Boltzmann equation. In the case
of a gas of inelastic hard spheres (IHS), it reads \cite{GS95,BDS97,vNE01}
\begin{equation}
\left( \partial _{t}+{\bf v\cdot \nabla }\right)
f=J^{(\alpha)}[f,f],
\label{3.1}
\end{equation}
where   $J^{(\alpha)}[f,f]$ is the Boltzmann collision operator
\begin{eqnarray}
J^{(\alpha)}[f,f]=\sigma ^{2}\int \dd{\bf v}_{1}\int
\dd\widehat{\bm{\sigma}}\,\Theta (
{\bf g}\cdot \widehat{\bm{\sigma}})({\bf g}\cdot \widehat{\bm{\sigma}})\nonumber\\
\times\left[ \alpha ^{-2}f({\bf v}'')f({\bf v} _{1}'')-f({\bf
v})f({\bf v}_{1})\right] ,
\label{3.2}
\end{eqnarray}
the explicit dependence of $f$ on ${\bf r}$ and $t$ having been
omitted. In Eq.\ (\ref{3.2}), $\sigma$ is the diameter of a sphere,
$\Theta $ is the Heaviside step function, $\widehat{\bm{\sigma}}$ is
a unit vector directed along the centers of the two colliding
spheres at contact, $\mathbf{g}=\mathbf{v}-\mathbf{v}_1$ is the
relative velocity, and $\alpha$ is the coefficient of normal
restitution.
 The pre-collisional or
restituting velocities ${\bf v}''$ and ${\bf v}_1''$ are given by
\begin{equation}
{\bf v}''={\bf v}-\frac{1+\alpha }{2\alpha}({\bf g}\cdot
\widehat{\bm{\sigma}})\widehat{\bm{\sigma}},\quad {\bf v}_{1}''={\bf
v}_{1}+\frac{1+\alpha }{2\alpha}({\bf g}\cdot \widehat{\bm{\sigma}
})\widehat{\bm{\sigma}},
\label{3.3}
\end{equation}
while the direct collision rule is
\begin{equation}
{\bf v}'={\bf v}-\frac{1+\alpha }{2}({\bf g}\cdot
\widehat{\bm{\sigma}})\widehat{\bm{\sigma}},\quad {\bf v}_{1}'={\bf
v}_{1}+\frac{1+\alpha }{2}({\bf g}\cdot \widehat{\bm{\sigma}
})\widehat{\bm{\sigma}}.
\label{3.3bis}
\end{equation}

The inelasticity of collisions contributes to a decrease of the
granular temperature $T(t)$, i.e.,
\begin{equation}
\frac{m}{3n}\int \dd{\bf v} V^{2}
 J^{(\alpha)}[f,f]=
-\zeta T,  \label{3.4}
\end{equation}
where  $\zeta$ is the \textit{cooling rate}. By standard manipulations of the collision
operator, the cooling rate can be written as \cite{BDS97,BDS99}
\beq
\zeta =\frac{\tau^{-1}}{48}\left(\frac{m}{T}\right)^{3/2} \langle
V_{12}^3\rangle(1-\alpha ^{2}),
\label{3.5}
\eeq
where
\beq
\langle V_{12}^3\rangle =\frac{1}{n^2} \int \dd{\bf v}_1\int \dd{\bf
v}_{2}\,|{\bf v }_1-{\bf v}_{2}|^{3}f({\bf v}_1)f({\bf v}_{2})
\label{3.5bis}
\eeq
is the (local) average value of the cube of the relative speed  and
\beq
\tau=\frac{\lambda}{\sqrt{2T/m}}
\label{2.11}
\eeq
is a (local) characteristic time,
\beq
\lambda=\left(\sqrt{2}\pi n\sigma^2\right)^{-1}
\label{2.12}
\eeq
being the (local) mean free path. Note that the cooling rate $\zeta$ is a
\textit{nonlinear functional} of the distribution function $f$ through the
average $\langle V_{12}^3\rangle$ and  cannot be explicitly
evaluated without the knowledge of $f$. Nevertheless, a simple \textit{estimate}  is obtained
from Eq.\ (\ref{3.5bis}) by replacing the actual distribution
function $f$ by the \textit{local equilibrium} distribution
\begin{equation}
f_0(\mathbf{v})=n(m/2\pi T)^{3/2}\exp(-mV^2/2T).
\label{3.5.1}
\end{equation}
In that case,
\beq
\langle V_{12}^3\rangle\to \langle
V_{12}^3\rangle_0=\frac{32}{\sqrt{\pi}}\left(\frac{T}{m}\right)^{3/2}.
\label{n2.1}
\eeq
Insertion of this approximation  into Eq.\ (\ref{3.5}) yields the
\textit{local equilibrium} cooling rate
 \cite{BDS99,BDKS98}
\begin{equation}
\zeta_0=\frac{2\tau^{-1}}{3\sqrt{\pi}} (1-\alpha ^{2}). \label{3.14}
\end{equation}
 This local equilibrium estimate is still a functional of $f$, but
 only
through the local density and temperature, i.e., $\zeta_0\propto n T^{1/2}$. In addition, its
dependence on inelasticity is simply $\zeta_0\propto 1-\alpha^2$.

The characteristic time $\tau$ defined by Eqs.\ (\ref{2.11}) and (\ref{2.12}) is of the order of the (local
equilibrium) mean free time $\tau_{\text{mft}}$, namely
\beq
\tau_{\text{mft}}=\frac{\lambda}{\langle
V\rangle_0}=\frac{\sqrt{\pi}}{2}\tau,
\label{2.11.2}
\eeq
where in the last step we have taken into account that the mean
(peculiar) speed is $\langle V\rangle\to\langle
V\rangle_0=\sqrt{8T/\pi m}$ in the local equilibrium approximation.
It is also convenient to introduce a characteristic time
$\tau_\visc$ associated with the momentum transfer or
viscosity. Its expression is
\beq
\tau_\visc=\frac{\eta_0}{nT}=1.016\frac{5\sqrt{\pi}}{8}\tau,
\label{2.11.3}
\eeq
where $\eta_0=1.016\times 5\sqrt{mT/\pi}/16\sigma^2$ is the Navier--Stokes shear viscosity in the elastic
limit ($\alpha\to 1$) \cite{CC70}. Note that
$\tau_\visc/\tau\approx\tau/\tau_{\text{mft}}\approx 1.13$.

\subsection{(Frictional) elastic hard spheres}
Now we consider a dilute gas of elastic hard spheres (EHS) of the
same mass $m$ as the IHS but with a diameter
$\sigma'=[\beta(\alpha)]^{1/2}\sigma$ smaller than that of the IHS.
As a consequence,  the
characteristic time $\tau'$ and the mean free path $\lambda'$ of the EHS gas
are \cite{note1}
\beq
\tau'=\frac{\lambda'}{\sqrt{2T/m}},\quad
\lambda'=\frac{1}{\beta}\left(\sqrt{2}n\pi\sigma^2\right)^{-1}.
\label{lambda'}
\eeq
Furthermore, we assume that the EHS are under the influence of a
drag force $\mathbf{F}_{\text{drag}}=-m\gamma(\alpha)\mathbf{V}$ with a
friction constant $\gamma(\alpha)=\frac{1}{2}\zeta_0(\alpha)$, where
$\zeta_0(\alpha)$ is given by Eq.\ (\ref{3.14}). Therefore, the
Boltzmann equation for the (frictional) EHS gas is
\begin{equation}
\left( \partial _{t}+{\bf v\cdot \nabla }-\frac{\zeta_0
}{2}\frac{\partial }{\partial {\bf v}}\cdot {\bf V}\right) f=\beta
J^{(1)}[f,f],
\label{n2.3}
\end{equation}
where the elastic collision operator $J^{(1)}[f,f]$ is given by Eq.\
(\ref{3.2}) by setting $\alpha=1$ both explicitly and in the
collision rule (\ref{3.3}), but keeping the factor $\sigma^2$.
Henceforth, when referring to the EHS system, we will always
understand
that the particles are frictional, in the sense that the external
force $\mathbf{F}_{\text{drag}}$ is acting on them \cite{note0}.

Friction produces in the EHS gas a cooling effect characterized by
the rate $\zeta_0(\alpha)$. This is intended to mimic the cooling
effect in the IHS gas due to the collisional inelasticity, which is
characterized by the  rate $\zeta(\alpha)$ given by Eq.\
(\ref{3.5}). Both cooling rates are quantitatively close each other
as long as the density and temperature are similar in both systems
and  (\ref{n2.1}) is a good approximation.
In principle, we could consider a friction constant $\gamma=\frac{1}{2}\zeta_0\langle V_{12}^3\rangle/\langle
V_{12}^3\rangle_0$, so that the cooling rate of the EHS would be the same
functional of $f$ as in the IHS case. However, this would complicate
excessively the EHS model without a correlated gain in accuracy, as
we will see in Sec.\ \ref{sec5C}.

The parameter
$\beta(\alpha)=(\sigma'/\sigma)^2$ can be adjusted to optimize the agreement
between the physically most relevant
integrals involving $J^{(\alpha)}[f,f]$ and $\beta
J^{(1)}[f,f]+\frac{1}{2}\zeta_0\partial_\mathbf{v}\cdot (\mathbf{V}f)$.
The simplest choice is \cite{SA04}
\beq
\beta=\frac{1+\alpha}{2}.
\label{n3.12}
\eeq

Of course, the IHS and EHS gases described by the Boltzmann
equations (\ref{3.1}) and (\ref{n2.3}), respectively, are
intrinsically different. However, it might be possible that the main
transport properties, which are measured by low-velocity moments
such as in Eqs.\ (\ref{b1})--(\ref{c1}), are similar in both systems. As said in Sec.\ \ref{sec1}, one of the goals of this paper is to
check this expectation in the case of the uniform shear flow
\cite{AS04}.

\section{Uniform shear flow \label{sec3}}
The uniform (or simple) shear flow (USF) is perhaps the nonequilibrium
state most widely studied, both for granular
\cite{SGD03,LSJC84,CG86,WB86,JR88,C89,C90,S92,HS92,SK94,LB94,GT96,SGN96,C97,BRM97,CR98,MGSB99,K00,F00,C01,AH01,CH02,G02,MG02,BGS02,GM03,AL03,L04,MGAS04}
and conventional \cite{GS03} gases. In this state the gas is
enclosed between two infinite parallel planes located at $y=-L/2$
and $y=+L/2$ and in relative motion with velocities $-U/2$ and
$+U/2$, respectively, along the $x$-direction. The planes do not
represent realistic bounding walls, in contrast to what happens in the true
Couette flow \cite{TTMGSD01}. Instead, the planes represent virtual
boundaries where the Lees--Edwards boundary conditions are applied
\cite{LE72,DSBR86}: every time a particle crosses one of the planes
with a given velocity $\mathbf{v}$, it is reentered at once through
the opposite plane with a velocity $\mathbf{v}'$ such that the
relative velocity with respect to the plane is preserved, i.e.,
$\mathbf{v}'=\mathbf{v}-U\widehat{\mathbf{x}}$ if the particle is
reentered through the bottom plate and
$\mathbf{v}'=\mathbf{v}+U\widehat{\mathbf{x}}$ if it is reentered
through the top plate. In terms of the velocity distribution
function, these generalized periodic boundary conditions read
\begin{equation}
f(y=\pm L/2, {\bf v};t)=f(y=\mp L/2, {\bf v}\mp  U \widehat{\bf
x};t).
\label{3.2:n4}
\end{equation}

This process injects energy into the system. Suppose a particle with
a velocity $\mathbf{v}$ crosses the top plane (i.e., $v_y>0$); it is
then transferred to the bottom plane with a new velocity
$\mathbf{v}'=\mathbf{v}-U \widehat{\bf x}$. The change in kinetic
energy is therefore proportional to ${v'}^2-v^2=2U(U/2-v_x)$, which
is positive (negative) if $v_x<U/2$ ($v_x>U/2)$. Thus, some
particles gain energy while other particles lose energy through the
boundary conditions. On the other hand, the shearing motion tends to
produce a negative shear stress ($P_{xy}<0$), so that particles
moving upward near the top wall  preferentially have $v_x-U/2<0$.
Therefore, ${v'}^2-v^2>0$ on average. This \textit{viscous heating}
effect tends to increase the temperature, in opposition to the
cooling effect due to the inelasticity of collisions (in the IHS
case) or to the drag force (in the EHS case).

Starting from a given initial condition
$f(\mathbf{r},\mathbf{v};0)$, and after a certain transient regime,
the system is expected to reach a nonequilibrium steady state where
the above heating and cooling effects cancel each other. By symmetry
reasons, this steady state is characterized by uniform density and
temperature, and by a \textit{linear} velocity profile
$\mathbf{u}(\mathbf{r})=a y\widehat{\mathbf{x}}$, where $a=U/L$
represents the imposed shear rate. More generally, the steady
distribution function becomes \textit{uniform} when the velocities
of the particles are referred to the Lagrangian frame of reference
co-moving with the flow velocity \cite{DSBR86}:
\beq
f(\mathbf{r}, \mathbf{v})\to f(\mathbf{V}),\quad
\mathbf{V}=\mathbf{v}-a y\widehat{\mathbf{x}}.
\label{III.1}
\eeq
 If the initial distribution function $f(\mathbf{r},\mathbf{v};0)$ depends on space only
through the coordinate $y$ normal to the plates, this property is
maintained by the Boltzmann equations (\ref{3.1}) and (\ref{n2.3}),
so that one can make
\beq \mathbf{v}\cdot\nabla\to
v_y\frac{\partial}{\partial y}.
\label{III.2}
\eeq
Furthermore, if $f(\mathbf{r},\mathbf{v};0)$ is consistent with the
symmetry property (\ref{III.1}), again this is maintained by the
Boltzmann equations (\ref{3.1}) and (\ref{n2.3}), what implies that
\beq \mathbf{v}\cdot\nabla\to -a
V_y\frac{\partial}{\partial V_x}.
\label{III.3}
\eeq
In this latter situation, Eqs.\ (\ref{3.1}) and (\ref{n2.3}) become
spatially homogeneous equations since, in agreement with Eq.\ (\ref{III.3}), the effect of convection is
played by the non-conservative \textit{inertial} force
$\mathbf{F}_{\text{shear}}=-m a V_y \widehat{\mathbf{x}}$. In what
follows, we will refer to the transient solution of Eqs.\
(\ref{3.1}) and (\ref{n2.3}) with the replacement
(\ref{III.3}) as the \textit{homogeneous}  transient
problem.

On the other hand, if the initial distribution depends spatially on
$y$ only but does not become uniform under the transformation
(\ref{III.1}), then the replacement (\ref{III.2}) is valid but (\ref{III.3}) is not. In
that case, Eqs.\ (\ref{3.1}) and (\ref{n2.3}) cannot be made
equivalent to uniform equations and their transient solutions define
the \textit{inhomogeneous} transient problem.

For the inhomogeneous transient problem we will consider two different initial states. The first one is the
\textit{equilibrium}  state
\beq
f(\mathbf{r},\mathbf{v};0)=\overline{n}\left(\frac{m}{2\pi
T^0}\right)^{3/2} \exp\left(-\frac{mv^2}{2T^0}\right),
\label{III.4}
\eeq
where $\overline{n}$ and $T^0$ are constants \cite{note2}. The gas is initially at
rest (in the laboratory or Eulerian frame), but almost immediately
the Lees--Edwards boundary conditions produce fluid motion near the
walls, this motion being subsequently propagated to the rest of the
system through  free streaming and collisions. Eventually the
flow velocity reaches the linear profile $u_x(y)=ay$. The transient
period
from $u_x=0$ to $u_x=ay$ induces inhomogeneities in the density
 and temperature  profiles, even though these quantities are initially uniform.
As a second choice for the initial state we will take the
distribution
\beq
f(\mathbf{r},\mathbf{v};0)=\frac{n^0(y)}{4\pi {V^0}^2}\delta\left(|\mathbf{v}-\mathbf{u}^0(y)|-V^0\right),
\label{new1}
\eeq
where the initial density and velocity fields are
\beq
n^0(y)=\overline{n}\left(1+\frac{1}{2}\sin \frac{2\pi y}{L}\right),
\label{new2}
\eeq
\beq
\mathbf{u}^0(y)=U\left(\cos\frac{\pi
y}{L}-\frac{2}{\pi}\right)\widehat{\mathbf{x}},
\label{new3}
\eeq
respectively, while the initial temperature $T^0=m{V^0}^2/3$ is uniform. The
initial state (\ref{new1}) is very different from (\ref{III.4}). Now, all the particles have the same magnitude
$V^0$ of the peculiar velocity. In addition, the initial density and flow
velocity fields have the opposite symmetries to the ones imposed by
the boundary conditions:
$n^0(-y)-\overline{n}=-\left[n^0(y)-\overline{n}\right]$ and
$\mathbf{u}^0(-y)=\mathbf{u}^0(y)$. Therefore, high gradients are expected during the period of time before the boundary
conditions establish a linear velocity profile and uniform density
and temperature.

Regardless of the initial preparation of the system, conservation of the total
number of particles implies that the average density coincides  for all times with
the initial value $\overline{n}$, i.e.,
\beq
\overline{n}=\frac{1}{L}\int_{-L/2}^{L/2}\dd y\, n(y,t).
\label{III.5}
\eeq
 However, the average temperature
\beq
\overline{T}(t)=\frac{1}{\overline{n}L}\int_{-L/2}^{L/2}\dd y\, n(y,t)T(y,t)
\label{III.6}
\eeq
changes in time during the transient regime as a consequence of the
competition between the dissipative cooling and the viscous heating.

A
physically motivated way of measuring time is through the
accumulated number of collisions per particle $s(t)$ from the initial state
to time $t$. In the local equilibrium approximation,
$s(t)=s_0(t)$, where
\beq
s_0(t)=\frac{1}{2}\frac{2}{\sqrt{\pi}}\frac{1}{\overline{n}L}\int_0^t\dd
t'\int_{-L/2}^{L/2}\dd y\,\frac{n(y,t')}{\tau(y,t')}.
\label{III.8}
\eeq
Here, the local characteristic time $\tau(y,t)$ is given by Eqs.\
(\ref{2.11}) and (\ref{2.12}). The factor  $\frac{1}{2}$ takes into account
that two particles are involved in each collision, while the factor $2/\sqrt{\pi}$ accounts for the
relation $\tau_{\text{mfp}}/\tau=\sqrt{\pi}/2$ [cf.\ Eq.\
(\ref{2.11.2})]. Note that
$s_0(t)$ is the (local equilibrium) number of collisions per
particle of the IHS gas. The equivalent quantity  for
the EHS gas, $s_0'(t)$, is obtained from Eq.\ (\ref{III.8}) by using the
corresponding local characteristic time $\tau'(y,t)$ defined by Eq.\
(\ref{lambda'}), instead of $\tau(y,t)$. In principle, $s_0'(t)\neq \beta s_0(t)$ and, more
generally, $s'(t)\neq \beta s(t)$, unless the density and
temperature profiles, and their history, are the same in both systems \cite{note1}.
We will come back to this point in Sec.\ \ref{sec5B}.

In the analysis of the homogeneous transient problem, we will start
from a \textit{local equilibrium} initial state
\beq
f(\mathbf{r},\mathbf{v};0)=\overline{n}\left(\frac{m}{2\pi
T^0}\right)^{3/2}
\exp\left[-\frac{m}{2T^0}\left(\mathbf{v}-ay\widehat{\mathbf{x}}\right)^2\right].
\label{III.7}
\eeq
In this case, the initial state is already uniform in the Lagrangian
frame [cf.\ Eq.\ (\ref{III.1})], so that the velocity field is kept linear, $u_x=ay$, the
density is constant, $n=\overline{n}$, and the temperature is uniform,
$T(y,t)=T(t)$. Consequently, Eq.\ (\ref{III.8}) becomes
\beq
s_0(t)=\frac{1}{\sqrt{\pi}}\int_0^t\frac{\dd t'}{\tau(t')},
\label{III.10}
\eeq
with a similar equation for $s_0'(t)$ in the EHS case.
Apart from the temperature $T(t)$ and the elements $P_{ij}(t)$ of the pressure
tensor, we will also evaluate in the homogeneous transient problem  the ratio
$\langle V_{12}^3\rangle/\langle V_{12}^3\rangle_0$, as well as the fourth and sixth
cumulants,
\beq
a_2=\frac{\langle V^4\rangle}{\langle V^4\rangle_0}-1,\quad a_3=-\frac{\langle V^6\rangle}{\langle V^6\rangle_0}+1+3a_2,
\label{III.19}
\eeq
where $\langle V^4\rangle_0={15}(T/m)^2$ and $\langle
V^6\rangle_0={105}(T/m)^3$. We recall that the quantity $\langle
V_{12}^3\rangle$ is directly related to the cooling rate of the IHS gas via
Eq.\ (\ref{3.5}), while the cumulants are measures of the deviation of the energy distribution from the Maxwellian.
Those deviations will also be monitored through the ratios
\beq
R_\ell(t)=\frac{\int_{W_\ell}^{W_{\ell+1}}dV\,V^2\int
d\widehat{\mathbf{V}}f(\mathbf{V},t)}{\int_{W_\ell}^{W_{\ell+1}}dV\,V^2\int
d\widehat{\mathbf{V}}f_0(\mathbf{V},t)},\quad \ell=0,1,2,3.
\label{new4}
\eeq
$R_\ell(t)$ is  the fraction of particles that at time
$t$ move with a  speed between $W_\ell$ and $W_{\ell+1}$, divided by the same fraction in local equilibrium.
We have taken for the integration limits the  values $W_\ell= C_\ell \sqrt{2T(t)/m}$ with $C_0=0$, $C_1=1$, $C_2=2$,
$C_3=3$, and $C_4=\infty$.

 In either transient problem, the final steady-state temperature $T_\s$ is smaller or larger than the
 initial value $T^0$ depending on whether initially the dissipative cooling
 dominates or is dominated by the viscous heating, respectively. By dimensional analysis,  $T_\s$ is proportional to
 $a^2$ times a certain function of $\alpha$, being independent of $T^0$. Stated differently,
 the steady state  is such that when the steady-state characteristic time
 $\tau_\s\propto T_\s^{1/2}$ is
 nondimensionalized with the constant shear rate  $a$, then it
 becomes a function of $\alpha$ only.
More generally, the \textit{reduced} steady-state velocity
distribution function
\beq
f_\s^*(\mathbf{C})=\frac{1}{n}\left(\frac{2T_\s}{m}\right)^{3/2}f_\s(\mathbf{V}),
\quad \mathbf{C}=\frac{\mathbf{V}}{\sqrt{2T_\s/m}},
\label{III.12}
\eeq
depends on $\alpha$ but is independent of the shear rate $a$ and the
initial preparation of the system. Owing to the symmetry properties
of the USF,
\beq
f_\s^*(C_x,C_y,C_z)=f_\s^*(C_x,C_y,-C_z)=f_\s^*(-C_x,-C_y,C_z).
\label{III.13}
\eeq
Since $f_\s^*(\mathbf{C})$ depends on the three velocity components
in a non-trivial way, it is difficult to visualize it, so that it is convenient to consider the following
\textit{marginal} distributions:
\beq
g_x^{(\pm)}(C_x)=\int_{-\infty}^\infty\dd
C_z\int_{-\infty}^\infty\dd C_y\, \Theta(\pm C_y)f_\s^*(\mathbf{C}),
\label{III.15}
\eeq
\beq
g_y^{(\pm)}(C_y)=\int_{-\infty}^\infty\dd
C_z\int_{-\infty}^\infty\dd C_x\, \Theta(\pm C_x)f_\s^*(\mathbf{C}),
\label{III.16}
\eeq
\beq
F(C)=C^2\int\dd \widehat{\mathbf{C}}\,f_\s^*(\mathbf{C}).
\label{III.14}
\eeq
 The function
$g_x^{(+)}(C_x)$ is the distribution of the $x$-component of the
velocity of those particles moving upward (i.e., with $C_y>0$). The functions
$g_x^{(-)}(C_x)$, $g_y^{(+)}(C_y)$, and $g_y^{(-)}(C_y)$ have a
similar meaning.
The symmetry properties (\ref{III.13}) imply that
\beq
g_x^{(+)}(C_x)=g_x^{(-)}(-C_x), \quad
g_y^{(+)}(C_y)=g_y^{(-)}(-C_y).
\label{III.17}
\eeq
While the functions (\ref{III.15}) and (\ref{III.16}) provide information about the anisotropy of the state,
$F(C)$ is the probability distribution function of the magnitude of
the peculiar velocity (in units of the thermal speed), regardless of its orientation.

\section{Monte Carlo simulations\label{sec4}}
We have solved numerically the Boltzmann equation (\ref{3.1}) for
the IHS system and the Boltzmann equation (\ref{n2.3}) for the EHS
system by means of the Direct Simulation Monte Carlo (DSMC) method
\cite{B94,AG97,MS96}. For the sake of completeness, we give below some details about the
application of this method to our problem.

\subsection{Inhomogeneous transient problem\label{sec4.A}}
 The USF has been implemented in the
inhomogeneous transient problem by applying the Lees--Edwards
boundary conditions (\ref{3.2:n4}), using the form (\ref{III.2}) for
the convection operator, and starting from the initial distribution
(\ref{III.4}) or (\ref{new1}). The separation between the boundaries has been taken
as $L=2.5\lambda^0$ and the shear rate has been fixed at
$a\tau^0=4$, where
\beq
\lambda^0=\left(\sqrt{2}\pi\overline{n}\sigma^2\right)^{-1},\quad
\tau^0=\frac{\lambda^0}{\sqrt{2T^0/m}}
\label{IV.1}
\eeq
are  the initial (global) mean free path and characteristic time,
respectively, of the IHS gas. The coefficient of restitution for the
IHS has been taken as $\alpha=0.9$. This same value has been taken in
the EHS case for the friction constant $\gamma=\frac{1}{2}\zeta_0$ and
the factor $\beta$, as given by Eqs.\ (\ref{3.14}) and
(\ref{n3.12}), respectively.

According to the DSMC method \cite{B94,AG97}, the system is split into $M$
layers of width $\delta y=L/M$. The velocity distribution function
is represented by the positions $\{y_i(t)\}$ and velocities
$\{\mathbf{v}_i(t)\}$ of a set of $N$ simulated particles:
\beq
f(y,\mathbf{v};t)\to \frac{1}{A}\sum_{i=1}^N
\delta(y-y_i(t))\delta(\mathbf{v}-\mathbf{v}_i(t)),
\label{IV.4}
\eeq
where $A\equiv (N/L)/\overline{n}$  is a constant formally representing the area of a section of
the system normal to the $y$-axis, so that  $A\delta y$ represents the volume
of a layer. The number of particles inside a given layer
$I=1,\ldots,M$ is
\beq
N_I(t)=\sum_{i=1}^N\Theta_I(y_i(t)),
\label{IV.5}
\eeq
where $\Theta_I(y)$ is the characteristic function of layer $I$,
i.e., $\Theta_I(y)=1$ if $y$ belongs in $I$ and $\Theta_I(y)=0$
otherwise. The (coarse-grained) number density, mean velocity,
temperature, and pressure tensor of layer $I$ are
\beq
n_I(t)=\frac{N_I(t)}{A\delta y},
\label{IV.6}
\eeq
\beq
\mathbf{u}_I(t)=\frac{1}{N_I(t)}\sum_{i=1}^N\Theta_I(y_i(t))\mathbf{v}_i(t),
\label{IV.7}
\eeq
\beq
T_I(t)=\frac{m}{3N_I(t)}\sum_{i=1}^N\Theta_I(y_i(t))\left[\mathbf{v}_i(t)-\mathbf{u}_I(t)\right]^2,
\label{IV.8}
\eeq
\beq
\mathsf{P}_I(t)=\frac{m}{A\delta
y}\sum_{i=1}^N\Theta_I(y_i(t))\left[\mathbf{v}_i(t)-\mathbf{u}_I(t)\right]\left[\mathbf{v}_i(t)-\mathbf{u}_I(t)\right].
\label{IV.9}
\eeq
The average temperature and pressure tensor along the system are
given by
\beq
\overline{T}(t)=\frac{1}{N}\sum_{I=1}^M N_I(t) T_I(t),
\label{IV.10}
\eeq
\beq
\overline{\mathsf{P}}(t)=\frac{1}{M}\sum_{I=1}^M \mathsf{P}_I(t).
\label{IV.10bis}
\eeq

The positions  $\{y_i(t)\}$ and velocities $\{\mathbf{v}_i(t)\}$ of the particles are
updated from time $t$ to time $t+\delta t$ in two stages:
\begin{enumerate}
\item
\textit{Free streaming.} In this stage,
\beq
y_i(t+\delta t)=y_i(t)+ v_{iy}(t)\delta t.
\label{IV.2}
\eeq
If particle $i$ crosses the top wall, i.e., $y_i(t+\delta t)>L/2$,
then its position and velocity are redefined as
\beq
y_i(t+\delta t)\to y_i(t+\delta t)-L,\quad \mathbf{v}_i(t)\to
\mathbf{v}_i(t)-aL\widehat{\mathbf{x}}.
\label{IV.3}
\eeq
A similar action takes place if particle $i$ crosses the bottom
wall, i.e., $y_i(t+\delta t)<-L/2$.

In the case of IHS, the velocities are not modified during the free
streaming stage. In the case of EHS, however, the action of the
(local) friction force yields
\beq
\mathbf{v}_i(t+\delta
t)=\mathbf{u}_I(t)+\left[\mathbf{v}_i(t)-\mathbf{u}_I(t)\right]e^{-\zeta_I(t)\delta
t/2}.
\label{IV.11}
\eeq
Here, $I$ is the layer where particle $i$ sits at time $t$ and
$\zeta_I(t)\propto n_I(t)\sqrt{T_I(t)}(1-\alpha^2)$ is the
coarse-grained version of the cooling rate $\zeta_0$ defined by
Eq.\  (\ref{3.14}).

\item
\textit{Collision stage.} In this stage, a number
\begin{subequations}
\beq
\mathcal{N}_I=\frac{N_I^2}{2\sqrt{2}N/M}\frac{w_I \delta
t}{\lambda^0} \quad \text{(IHS)},
\label{IV.12a}
\eeq
\beq
\mathcal{N}_I'=\beta(\alpha)\frac{N_I^2}{2\sqrt{2}N/M}\frac{w_I
\delta t}{\lambda^0}\quad \text{(EHS)}
\label{IV.12b}
\eeq
\label{IV.12}
\end{subequations}
of candidate pairs are randomly selected for each layer $I$. In
Eqs.\ (\ref{IV.12}), $\lambda^0$ is given by Eq.\ (\ref{IV.1}) and
$w_I\propto \sqrt{T_I(t)}$ is an upper estimate of the maximum
relative speed in layer $I$. The collision between  each candidate
pair $ij$ is accepted with a probability equal to $v_{ij}/w_I$,
where $v_{ij}$ is the relative speed. If the collision is accepted,
a direction $\widehat{\bm{\sigma}}$ is chosen at random with
equiprobability and the velocities $(\mathbf{v}_i, \mathbf{v}_j)$
are replaced by $(\mathbf{v}_i', \mathbf{v}_j')$, according to the
collision rule (\ref{3.3bis}) with $\alpha< 1$ (IHS) or $\alpha=1$
(EHS).

\end{enumerate}

 The numerical values for the ``technical''
parameters are as follows. The layer thickness is $\delta
y=0.05\lambda^0$ (i.e., the number of layers is $M=50$), the time
step is $\delta t=10^{-3}\tau^0\sqrt{T^0/\overline{T}}$  (so that it
changes in time as the global characteristic time $\overline{\tau}\propto
1/\sqrt{\overline{T}}$ does), and the total number of particles is
$N=10^4$. The hydrodynamic quantities (density, flow velocity,
temperature, and pressure tensor) are updated every 4 time
steps and recorded every 160 time steps. Moreover, in order to improve the statistics, all the
quantities are further averaged over three independent realizations
of the system.

\subsection{Homogeneous transient problem\label{sec4.B}}
In the homogeneous transient problem the USF is implemented by
working directly in the Lagrangian frame of reference [cf.\ Eq.\
(\ref{III.1})] and using (\ref{III.3}). Since the resulting
Boltzmann equation is uniform, only the (peculiar) velocities
$\{\mathbf{V}_i(t)\}$ of the $N$ particles need to be stored and
there is no need of splitting the system into cells or applying
boundary conditions. The velocity distribution function is described
by
\beq
f(\mathbf{V};t)\to \frac{1}{\Omega}\sum_{i=1}^N
\delta(\mathbf{V}-\mathbf{V}_i(t)),
\label{IV.4bis}
\eeq
where the constant $\Omega\equiv N/\overline{n}$ formally represents the volume of the system.
The temperature and the pressure tensor are evaluated as
\beq
{T}(t)=\frac{m}{3N}\sum_{i=1}^N{V}_i^2(t),
\label{IV.14}
\eeq
\beq
\mathsf{P}(t)=\frac{m}{\Omega}\sum_{i=1}^N
 \mathbf{V}_i(t)\mathbf{V}_i(t).
\label{IV.15}
\eeq
Analogously, the fourth and sixth cumulants of the distribution
function [cf.\ Eq.\ (\ref{III.19})] and the ratios $R_\ell$ [cf.\ Eq.\
(\ref{new4})] are computed as
\beq
a_2(t)=\frac{m^2}{15T^2(t)}\frac{1}{N}\sum_{i=1}^N V_i^4(t)-1,
\label{IV.16}
\eeq
\beq
a_3(t)=-\frac{m^3}{105T^3(t)}\frac{1}{N}\sum_{i=1}^N
V_i^6(t)+1+3a_2(t),
\label{IV.17}
\eeq
\beqa
R_\ell(t)&=&\frac{N^{-1}}{\mathcal{P}(C_{\ell+1})-\mathcal{P}(C_\ell)}\sum_{i=1}^N
\Theta\left({V_i(t)}-W_\ell(t)\right)\nn
&&\times
\Theta\left(W_{\ell+1}(t)-{V_i(t)}\right),
\label{new6}
\eeqa
where $\mathcal{P}(x)=\text{erf}(x)-2xe^{-x^2}/\sqrt{\pi}$,
$\text{erf}(x)$ being the error function.
In order to evaluate the average of the cube of the relative speed, a
sample of $N_p$ pairs is randomly chosen out of the total number
$N(N-1)/2$ of pairs, so that
\beq
\langle
V_{12}^3\rangle(t)=\frac{1}{N_p}\sum_{ij}|\mathbf{V}_i(t)-\mathbf{V}_j(t)|^3.
\label{IV.18}
\eeq

The velocity update $\{\mathbf{V}_i(t)\}\to \{\mathbf{V}_i(t+\delta
t)\}$ takes place again in two stages. In the free streaming stage
for IHS, only the $x$-component of the velocities change according
to the inertial force $\mathbf{F}_{\text{shear}}=-m a V_y
\widehat{\mathbf{x}}$:
\beq
V_{ix}(t+\delta t)=V_{ix}(t)-V_{iy}(t)a\delta t.
\label{IV.19}
\eeq
On the other hand, for EHS we have the drag force
$\mathbf{F}_{\text{drag}}=-m(\zeta_0/2)\mathbf{V}$ in addition to
$\mathbf{F}_{\text{shear}}$, so that
\beq
\begin{aligned}
V_{ix}(t+\delta t)=\left[V_{ix}(t)-V_{iy}(t)a\delta
t\right]e^{-\zeta_0(t)\delta t/2},&\\
V_{iy,z}(t+\delta t)=V_{iy,z}(t)e^{-\zeta_0(t)\delta t/2},&
\end{aligned}
\label{IV.20}
\eeq
where $\zeta_0(t)$ is given by Eq.\ (\ref{3.14}) with
$\tau(t)=\lambda^0/\sqrt{2T(t)/m}$.

The collision stage proceeds essentially as in the inhomogeneous
case, except that formally the number of layers is $M=1$. Therefore,
a number
\begin{subequations}
\beq
\mathcal{N}=\frac{N}{2\sqrt{2}}\frac{w \delta t}{\lambda^0}\quad
\text{(IHS)},
\label{IV.13a}
\eeq
\beq
\mathcal{N}'=\beta(\alpha)\frac{N}{2\sqrt{2}}\frac{w \delta
t}{\lambda^0}\quad \text{(EHS)}
\label{IV.13b}
\eeq
\label{IV.13}
\end{subequations}
of candidate pairs are randomly selected out of
the total number of pairs in the system, $w\propto \sqrt{T(t)}$
being an upper estimate of the maximum value of the relative speeds
$\{V_{ij}\}$ in the whole system.

In the simulations of the homogeneous transient problem we have
considered two values of the shear rate ($a\tau^0=4$ and
$a\tau^0=0.1$) and ten values of the coefficient of restitution
($\alpha=0.5$--$0.95$ with a step $\Delta\alpha=0.05$), both for IHS
and EHS. This gives a total of forty different systems simulated.
However, in the discussion of the transient problem  we will mainly report
results corresponding to $\alpha=0.5$ and $\alpha=0.9$. Once a
steady state is reached, its properties are obtained by averaging
the  fluctuating simulation values over time. In the analysis of the
steady state all the values $\alpha=0.5$--$0.95$ will be considered.

The technical parameters of the homogeneous simulations are $\delta
t=10^{-3}\tau^0\sqrt{T^0/{T}}$ (the average quantities being updated
every 4 time steps and recorded every 160 time steps), $N=10^4$, and $N_p=2.5\times 10^4$.

\section{Results\label{sec5}}
In this Section we present the simulation results obtained for the
main physical quantities in the USF and compare the properties of
the genuine IHS gas with those of the  EHS gas.

\subsection{Inhomogeneous transient problem}
\begin{figure}
\includegraphics[width=1. \columnwidth]{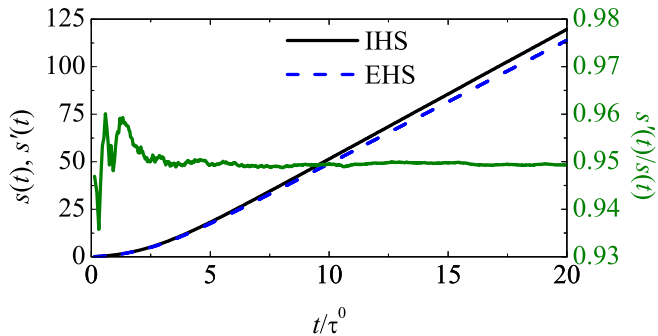}
 \caption{(Color online)
Accumulated number of collisions per particle as a function of time
for IHS [$s(t)$, smooth solid line] and EHS [$s'(t)$, smooth dashed
line] in the case $\alpha=0.9$, $a\tau^0=4$. The fluctuating
solid line represents the ratio $s'(t)/s(t)$, the corresponding scale being that of the right vertical axis.
The data have been obtained starting from the initial distribution function  (\protect\ref{III.4}).\label{inhom_coll}}
 \end{figure}
\begin{figure*}
\includegraphics[width=2. \columnwidth]{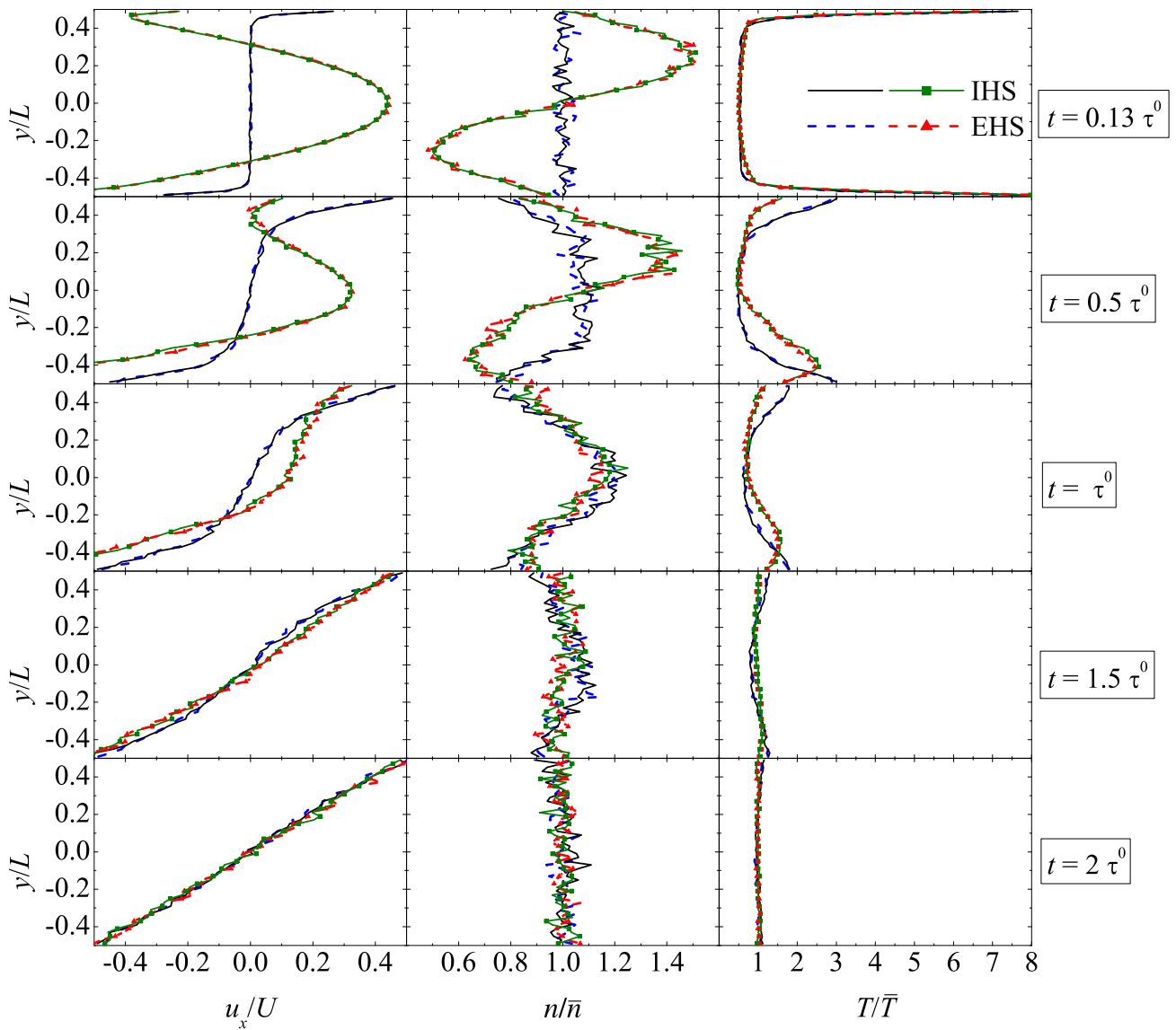}
 \caption{(Color online)
Hydrodynamic profiles for IHS (solid lines) and EHS (dashed lines)
in the case $\alpha=0.9$, $a\tau^0=4$, at times
$t/\tau^0=0.13$, $0.5$, $1$, $1.5$, and $2$.
The data have been obtained starting from the initial distribution function  (\protect\ref{III.4}) (curves without symbols) or
(\protect\ref{new1}) (curves with symbols).\label{profiles}}
 \end{figure*}
As said in Sec.\ \ref{sec4}, the initial distribution function  in
the inhomogeneous transient problem is either that of equilibrium, Eq.\ (\ref{III.4}),
or a strongly nonequilibrium one, Eq.\ (\ref{new1}).
We have
restricted ourselves to a coefficient of restitution $\alpha=0.9$
and a shear rate $a=4/\tau^0=(4/0.95)/{\tau^0}'$, where $\tau^0$ and
${\tau^0}'=\tau^0/\beta(\alpha)=\tau^0/0.95$ are the initial
(global)
characteristic times of the IHS and EHS gases, respectively.  The
values of $a$ and $\alpha$ are such that the viscous heating
initially prevails over the dissipative cooling (either collisional
or frictional) and so the temperature increases and the mean free
time decreases.

We will mainly monitor the temporal evolution of the
physical quantities by using an \textit{internal} clock, namely the
accumulated number of collisions per particle  $s(t)$ (IHS) and
$s'(t)$ (EHS), rather than the \textit{external} time $t$. The
quantities $s(t)$ and $s'(t)$ are computed directly by dividing the
total number of accepted collisions until time $t$ by the total
number of particles; in general, they slightly differ from the local
equilibrium values $s_0(t)$  and $s_0'(t)$ [cf.\ Eq.\ (\ref{III.8})].
Insofar as the velocity distribution function $f(y,\mathbf{v};t)$ is
similar in both systems, one can expect that  $s'(t)/\beta(\alpha)\simeq s(t)$ \cite{note1}. Figure \ref{inhom_coll} shows
$s(t)$ and $s'(t)$ as  functions of time in the case of the initial state (\ref{III.4}). The slopes of those curves
are directly related to the respective temperatures; both slopes
increase monotonically until becoming constant for $t/\tau^0\gtrsim
8$. As expected, the accumulated number of collisions in the EHS gas
up to any given time $t$
 is smaller than in the IHS
gas, i.e., $s'(t)<s(t)$.  Figure \ref{inhom_coll} also shows the
temporal evolution of the ratio $s'(t)/s(t)$. It fluctuates around
$\beta(\alpha)=0.95$ up to $t/\tau^0\simeq 4$ and stays very close
to 0.95 thereafter.
 This already provides an
indirect validation of the practical equivalence between the
profiles and history of the hydrodynamic quantities in both systems.

Figure \ref{profiles} shows the velocity, density, and temperature
profiles at times $t/\tau^0=0.13$, $0.5$, $1$, $1.5$, and $2.0$. The
corresponding accumulated numbers of collisions per particle
(for IHS) are
$s\simeq  0.08$, $0.4$, $1.1$, $2.1$,  and $3.4$ in the case of the initial distribution (\ref{III.4}),
 and $s\simeq  0.09$, $0.5$, $1.5$, $2.8$,  and $4.6$ in the case of the initial distribution (\ref{new1}).
 By time
$t/\tau^0=0.13$ only about 16--18\% of particles have collided, so that
the deviations from the initial profiles are essentially due to the boundary conditions.
As a consequence, at $t/\tau^0=0.13$  the flow velocity is almost the initial one everywhere except
in the layers adjacent to the walls; moreover, those layers have a  much larger
temperature than the bulk, while the number density is still
practically unchanged. As time advances, the more energetic particles
near the boundaries travel inside the system and transfer part of
their momentum and energy to the other particles by means of
collisions. This produces a stretching of the shape of the
velocity profile as well as a homogenization of temperature. The
layers adjacent to the walls are first partially depopulated in favor of the
central layers, but as the velocity profile becomes linear and the
temperature becomes uniform, so does the density. In summary, Fig.\
\ref{profiles} clearly shows that the hydrodynamic profiles freely evolve toward the
characteristic profiles of the USF, regardless of the initial preparation of the system.
It must ne noted that, although the USF is known to be unstable
with respect to excitations of sufficiently long wavelengths
\cite{S92,SK94,GT96,K00}, the size of the simulated systems ($L=2.5\lambda^0$) is
small enough to suppress such an instability.
As a matter of fact, a recent  analysis from kinetic theory \cite{G05}
shows that at $\alpha=0.9$ the instability does  not appear unless
$L\gtrsim 25\lambda^0$.

\begin{figure}
\includegraphics[width=1. \columnwidth]{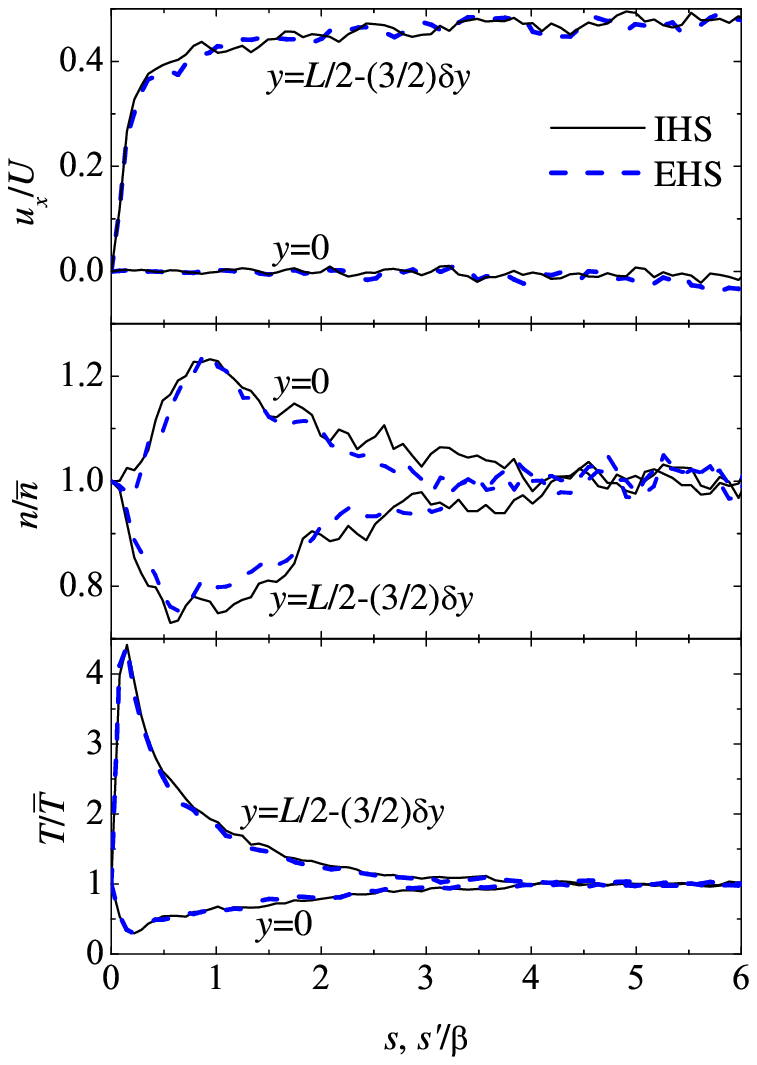}
 \caption{(Color online)
Evolution of the flow velocity, the density, and the temperature
around $y=0$ and $y=L/2-(3/2)\delta y$ for IHS (solid lines) and EHS (dashed lines) in
the case $\alpha=0.9$, $a\tau^0=4$. Time is measured by the
accumulated number of collisions per particle ($s$) in the case of
IHS and by the same quantity, but divided by $\beta$, ($s'/\beta$)
in the case of EHS. The data have been obtained starting from the initial distribution function  (\protect\ref{III.4}).\label{inhom_evol1}}
 \end{figure}
For the sake of clarity of the graphs, in the remainder of this Subsection we restrict
ourselves to present simulation results obtained
from the initial distribution (\ref{III.4}).
To monitor the time needed to establish a linear velocity profile
and uniform density and temperature, we have followed the evolution
of those quantities averaged over the four  central layers ($-2\delta
y\leq y\leq 2\delta y$) and over the three  top layers
($L/2-3\delta y\leq y\leq L/2$). The results are displayed in Fig.\
\ref{inhom_evol1}.  The flow velocity at $y\approx 0$ fluctuates around
zero, as expected by symmetry, whereas the velocity near the top
wall monotonically increases (except for fluctuations) toward the
wall velocity $U/2$. The maximum density difference appears after
about one collision per particle, while the maximum temperature
difference appears earlier at $s\simeq 0.15$. The characteristic
hydrodynamic profiles of USF, i.e., linear velocity and uniform
density and temperature, are reached approximately at $s=4$ (which
corresponds to a ``real'' time $t/\tau^0\simeq 2.2$). From this time
on, the evolution proceeds essentially as in the homogeneous
transient problem.

 \begin{figure}
\includegraphics[width=1. \columnwidth]{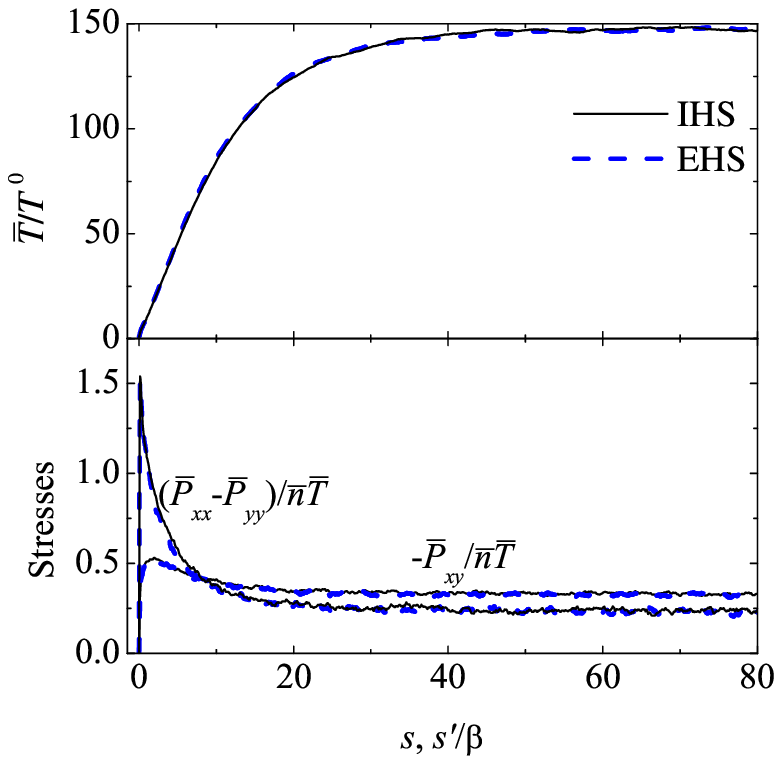}
 \caption{(Color online)
Evolution of $\overline{T}/T^0$ (top panel) and
$-\overline{P}_{xy}/\overline{n}\overline{T}$ and
$(\overline{P}_{xx}-\overline{P}_{yy})/\overline{n}\overline{T}$ (bottom panel) for IHS
(solid lines) and EHS (dashed lines) in the case $\alpha=0.9$,
$a\tau^0=4$. Time is measured by the accumulated number of
collisions per particle ($s$) in the case of IHS and by the same
quantity, but divided by $\beta$, ($s'/\beta$) in the case of
EHS. The data have been obtained starting from the initial distribution function  (\protect\ref{III.4}).\label{inhom_evol2}}
 \end{figure}
In Figs.\ \ref{profiles} and \ref{inhom_evol1} we have scaled the
local temperature $T(y,t)$ with respect to its average value
$\overline{T}(t)$ in order to focus on the transient period toward
uniformity. However, once the system becomes ``homogeneous'' [in the
sense of Eq.\ (\ref{III.1})] at $s\simeq 4$, the temperature
keeps evolving in time until the steady state is reached. This is
observed in the top panel of Fig.\ \ref{inhom_evol2}, which shows the
evolution of $\overline{T}/T^0$. We note that the total transient
period is much longer than the duration of the inhomogeneous state.
The temperature reaches a stationary value much larger than the
initial one ($\overline{T}_\s/T^0\simeq 147$) at $s\simeq 50$
($t/\tau^0\simeq 9.8$). Stationarity of temperature does not
necessarily imply that the steady state has been reached since, in
principle, the particles could redistribute their velocities along
time without altering the mean kinetic energy. A strong indication
that this is  not actually the case is provided by the bottom panel
of Fig.\ \ref{inhom_evol2}, which shows the evolution of the
(reduced) shear stress $-\overline{P}_{xy}/\overline{n}\overline{T}$ and the
(reduced) normal stress difference
$(\overline{P}_{xx}-\overline{P}_{yy})/\overline{n}\overline{T}$. These two quantities
reach stationary values $-\overline{P}_{xy,\s}/\overline{n}\overline{T}_\s\simeq
0.33$ and $(\overline{P}_{xx,\s}-\overline{P}_{yy,\s})/\overline{n}\overline{T}_\s\simeq
0.24$ after $s\simeq 40$ ($t/\tau^0\simeq 8.3$).

In the above  comments on Figs.\ \ref{profiles}--\ref{inhom_evol2}
we have focused on the physical features of the transient toward the
steady-state USF, without distinguishing between the IHS and EHS
systems. In fact, as Figs.\ \ref{profiles}--\ref{inhom_evol2} show,
the results concerning the hydrodynamic quantities and their fluxes
are practically identical in both systems, even when high gradients are transitorily present.
Therefore, the transport
properties of a gas of IHS can be satisfactorily mimicked by a gas
of EHS having an adequate diameter and subject to an adequate
friction force. However, although a coefficient of restitution
$\alpha=0.9$ is rather realistic, it is reasonable to expect that
this approximate equivalence IHS$\leftrightarrow$EHS deteriorates as
dissipation increases. To analyze this expectation we have
considered other values of $\alpha$ in the homogeneous  transient
problem and, especially, in the steady-state properties.

\subsection{Homogeneous transient problem\label{sec5B}}
As discussed in Sec.\ \ref{sec3}, the Boltzmann equation for USF allows
for solutions which are spatially uniform when the velocities are
referred to the (local) Lagrangian frame of reference [cf.\ Eq.\
(\ref{III.1})]. The simulation of the corresponding Boltzmann
equation by the DSMC method is much simpler than in the
inhomogeneous problem, as described in Sec.\ \ref{sec4.B}. In
these homogeneous simulations we have considered
$\alpha=0.5$--$0.95$ (with a step $\Delta \alpha=0.05$) and two
values of the shear rate, namely $a\tau^0=4$ and
$a\tau^0=0.1$. The former value  is large enough to make
viscous heating initially dominate over (inelastic or frictional)
cooling, even for $\alpha=0.5$, so that $T(t)>T^0$. Conversely,
$a\tau^0=0.1$ is small enough to produce the opposite effect,
$T(t)<T^0$, even for $\alpha=0.95$. On the other hand, at a given
value of $\alpha$, the intrinsic steady-state properties must be
independent of the value of the shear rate, and this will provide an
important indicator to determine whether the steady state has been
reached or not.

Although we have performed simulations for the ten values
$\alpha=0.5$--$0.95$, in this Subsection we will focus on two
values: $\alpha=0.9$ (moderately small dissipation) and $\alpha=0.5$
(large dissipation). In addition to the simulation data for IHS and
EHS, we will present the results from the solution of  an extension \cite{SA04,BDS99} of the Bhatnagar--Gross--Krook (BGK) model
\cite{BGK54}, which is inspired in the approximate equivalence
IHS$\leftrightarrow$EHS. The solution of this BGK-like model for the USF
problem is worked out in Refs.\ \cite{SA04,SGD03,BRM97}.

\begin{figure}
\includegraphics[width=1. \columnwidth]{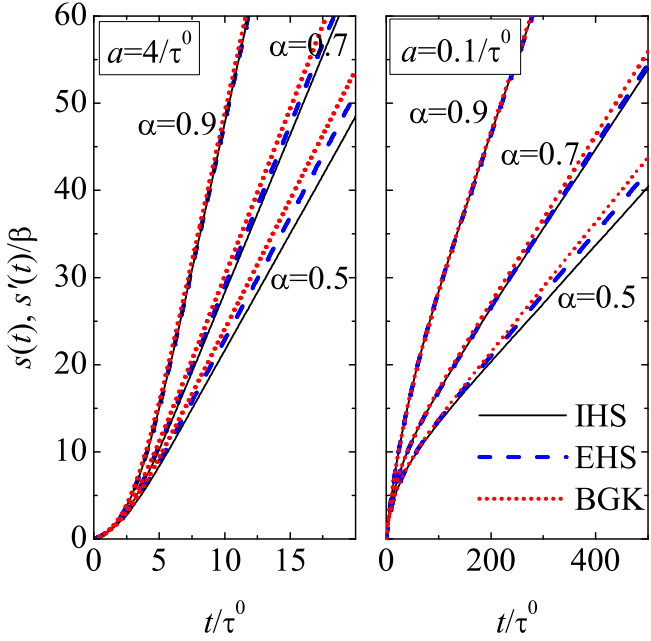}
 \caption{(Color online)
Accumulated number of collisions per particle as a function of time
 for IHS [$s(t)$, solid lines] and for EHS [$s'(t)$, dashed lines], in the latter case divided by
 $\beta=\frac{1}{2}(1+\alpha)$, in the cases $\alpha=0.5$, $0.7$, and
 $0.9$. The left panel corresponds to a shear rate $a\tau^0=4$, while the right panel corresponds to
 $a\tau^0=0.1$. The dotted lines are the predictions obtained
 from the solution of a BGK model. Note that
in the right panel the curves corresponding to IHS, EHS, and BGK at
$\alpha=0.9$ are practically indistinguishable.
The data have been obtained starting from the initial distribution function  (\protect\ref{III.7}).\label{hom_coll}}
 \end{figure}
 \begin{figure}
\includegraphics[width=1. \columnwidth]{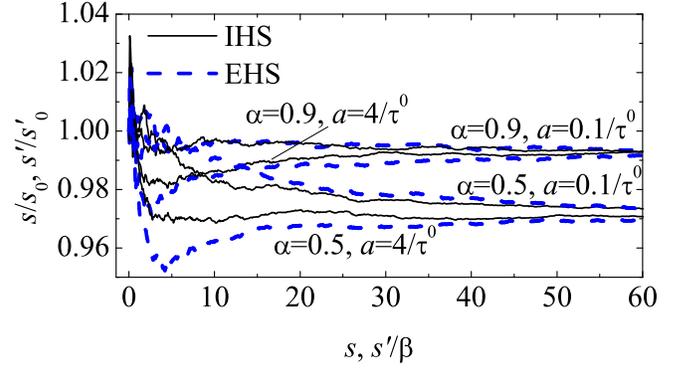}
 \caption{(Color online)
Evolution of the ratio between the actual number of collisions per
particle and the local equilibrium estimate for IHS (solid lines)
and EHS (dashed lines) in the cases  $\alpha=0.5, 0.9$ and
$a\tau^0=0.1, 4$.
The data have been obtained starting from the initial distribution function  (\protect\ref{III.7}).\label{hom_coll2}}
 \end{figure}
We begin by showing the accumulated number of collisions per particle as a
function of time in Fig.\ \ref{hom_coll}, where the case
$\alpha=0.7$ is also included. As said in connection with Fig.\
\ref{inhom_coll}, the slope of $s(t)$ and $s'(t)$ is proportional to
$\sqrt{T(t)}$. At a shear rate $a\tau^0=4$, viscous
heating dominates and so the temperature monotonically increases,
especially for $\alpha=0.9$; on the other hand, at
$a\tau^0=0.1$, dissipative cooling prevails and so the
temperature monotonically decreases, especially for $\alpha=0.5$.
The almost perfect agreement between $s(t)$ and $s'(t)/\beta$
for $\alpha=0.9$ is an indirect indication that $T(t)$ is
practically the same for the IHS gas and the EHS gas, as will be
confirmed later on. However, as the inelasticity increases, so does
the temperature difference in both systems, the EHS system having
 a slightly higher temperature than the IHS
system at any given time $t$. Despite its simplicity, the BGK model does quite good a
job, but it tends to overestimate both $s(t)$ and $s'(t)/\beta$. It must be
said that the BGK curves for $s(t)$ have actually been obtained from the local
equilibrium approximation (\ref{III.10}). As we will see, the BGK
temperature presents a very good agreement with the simulation
results for EHS, so that the small discrepancies between the DSMC
curves for $s'(t)/\beta$ and the BGK curves are essentially due to
the local equilibrium approximation $s(t)\to s_0(t)$ used in the latter.

To confirm
this point, we plot in Fig.\ \ref{hom_coll2} the ratio between the
actual number of accumulated collisions per particle and the local
equilibrium estimate obtained from Eq.\ (\ref{III.10}) by a
numerical integration using the actual values of temperature. Except
for a short initial period, $s(t)/s_0(t)$ and $s'(t)/s_0'(t)$ take
values smaller than 1 and tend to steady-state values practically
independent of the shear rate.
The instantaneous collision rate is proportional to the average
value of the relative speed $\langle V_{12}\rangle$, which in the
local equilibrium approximation is $\langle V_{12}\rangle\to \langle
V_{12}\rangle_0=\sqrt{2}\langle V\rangle_0=4\sqrt{T/\pi m}$. Thus, the fact that $s(t)<s_0(t)$ is
consistent with a nonequilibrium velocity distribution such that the
mean speed is smaller than the local equilibrium one, i.e., $\langle
V\rangle<\langle V\rangle_0$, this effect being more
noticeable as the dissipation increases. By definition,  $\langle
V^2\rangle=\langle V^2\rangle_0={3T/ m}$. Therefore, the inequality
$\langle V\rangle<\langle V\rangle_0$ indicates an underpopulation
of the  region of  moderately low velocities of the nonequilibrium distribution
function (with respect to the Maxwellian) that compensates for an
overpopulation of the high-velocity region, as  will be seen later.

  \begin{figure}
\includegraphics[width=1. \columnwidth]{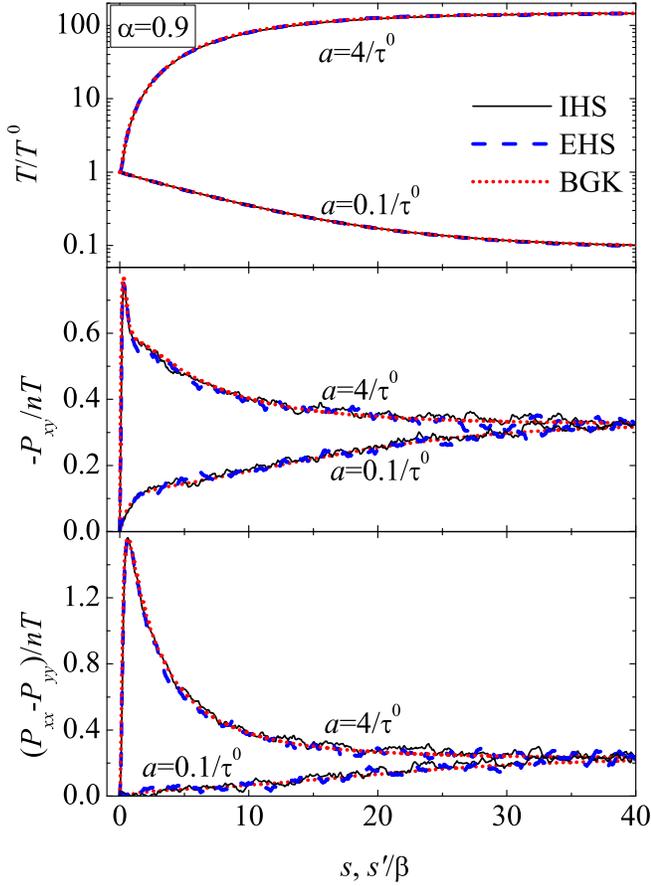}
 \caption{(Color online)
Evolution of  $T/T^0$,  $-P_{xy}/nT$, and  $(P_{xx}-P_{yy})/nT$ for
IHS (solid lines) and EHS (dashed lines) in the cases $\alpha=0.9$
with $a\tau^0=0.1$  and $a\tau^0=4$. The dotted lines are
the predictions obtained from the solution of a BGK model. Note that
in the top panel the curves corresponding to IHS, EHS, and BGK are
practically indistinguishable.
The data have been obtained starting from the initial distribution function  (\protect\ref{III.7}).\label{hom_hidro09}}
 \end{figure}
  \begin{figure}
\includegraphics[width=1. \columnwidth]{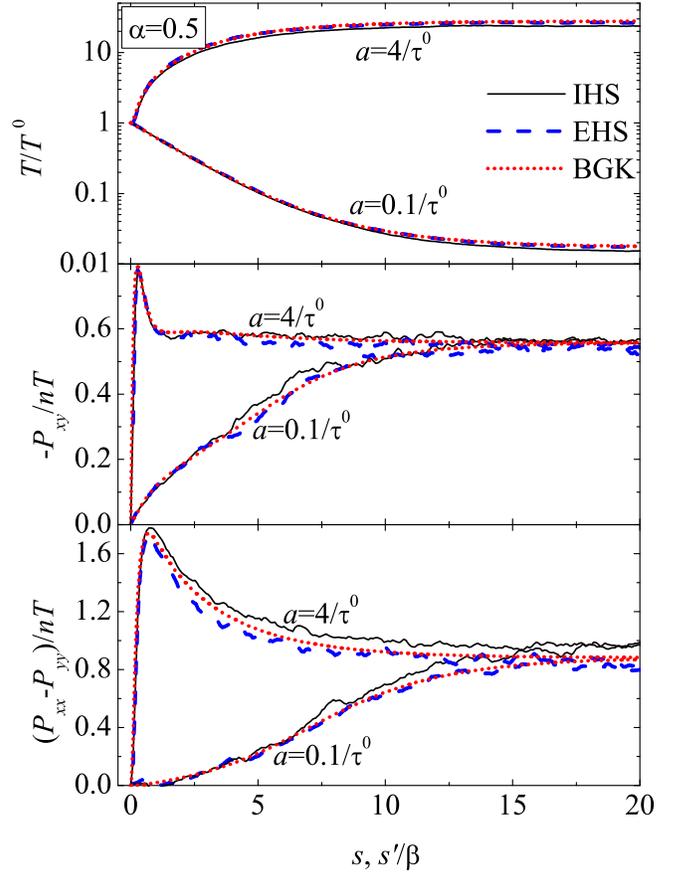}
 \caption{(Color online)
Evolution of  $T/T^0$,  $-P_{xy}/nT$, and  $(P_{xx}-P_{yy})/nT$ for
IHS (solid lines) and EHS (dashed lines) in the cases $\alpha=0.5$
with $a\tau^0=0.1$  and $a\tau^0=4$. The dotted lines are
the predictions obtained from the solution of a BGK model.
The data have been obtained starting from the initial distribution function  (\protect\ref{III.7}).
\label{hom_hidro05}}
 \end{figure}
The evolution of the relative temperature $T/T^0$, the reduced shear
stress $-P_{xy}/nT$, and the reduced normal stress difference
$(P_{xx}-P_{yy})/nT$ is displayed in Figs.\ \ref{hom_hidro09} and
\ref{hom_hidro05} for $\alpha=0.9$ and $\alpha=0.5$, respectively.
In the former case, an excellent agreement  between the
simulation results for both types of system exists. In addition, the
theoretical results obtained from the BGK model accurately describe
the behavior of the simulation data. In the case $\alpha=0.5$,
however, the EHS system tends to have a larger temperature than the
IHS system, the steady-state value being about 12\% larger in the
former case than in the latter. This is partially due to the fact
that the true cooling rate $\zeta$ of the IHS gas [cf.\ Eq.\
(\ref{3.5})] is larger than the local equilibrium value $\zeta_0$
[cf.\ Eq.\ (\ref{3.14})] imposed on the EHS gas, as we will see
later on. This also explains why the BGK model, which also makes use
of the approximation $\zeta\to\zeta_0$, predicts a temperature in
good agreement with the simulation data for EHS. Since the EHS
temperature is larger than the IHS one, the shear rate normalized
with the collision rate is smaller for EHS than for IHS and,
consequently, the distortion with respect to the Maxwellian, as
measured by the shear stress and, especially, by the normal stress
difference, is smaller in the former case than in the latter.
Comparison between Figs.\ \ref{hom_hidro09} and \ref{hom_hidro05}
shows that the duration of the transient period (as measured by the
number of collisions per particle) decreases as the inelasticity
increases: $s\simeq 40$ at $\alpha=0.9$ versus $s\simeq 20$ at
$\alpha=0.5$. However, when that duration is measured in real units,
it depends mainly on the shear rate and not on $\alpha$, namely
$t\simeq 9\tau^0$ for $a\tau^0=4$ and $t\simeq 160$--$200\tau^0$ for $a\tau^0=0.1$.

 \begin{figure}
\includegraphics[width=1. \columnwidth]{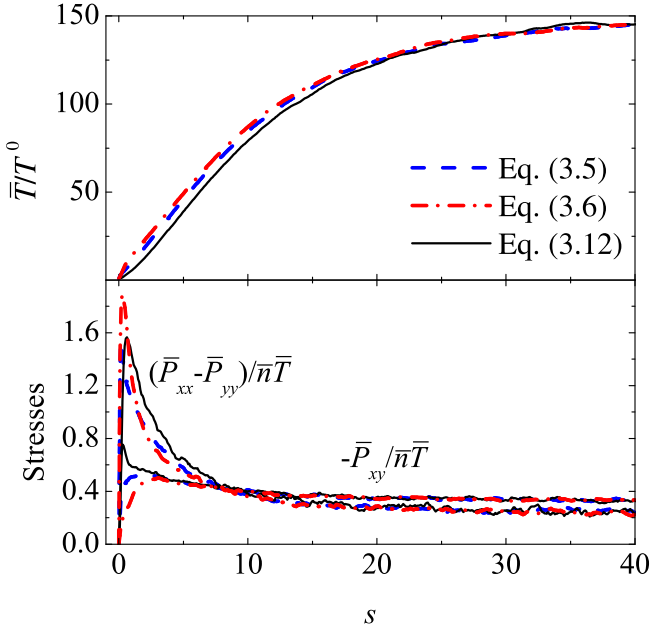}
 \caption{(Color online)
Evolution of $\overline{T}/T^0$ (top panel) and
$-\overline{P}_{xy}/\overline{n}\overline{T}$ and
$(\overline{P}_{xx}-\overline{P}_{yy})/\overline{n}\overline{T}$ (bottom panel) for IHS
in the case $\alpha=0.9$, $a\tau^0=4$. The dashed and dash-dot lines
correspond to the inhomogeneous transient problem with the initial conditions (\protect\ref{III.4})
and (\protect\ref{new1}), respectively, while  the solid
lines correspond to the homogeneous transient
problem with the initial conditions (\protect\ref{III.7}).\label{inhom_hom_evol}}
 \end{figure}
It is interesting to note the similarity between the curves in Fig.\
\ref{inhom_evol2} and those corresponding to $a\tau^0=4$ in
Fig.\ \ref{hom_hidro09}. This is made clear in Fig.\
\ref{inhom_hom_evol}, where the evolution of the global quantities in the IHS
case for the homogeneous and inhomogeneous transient problems are shown.
In the three situations considered, the initial shear stress
and normal stress differences are zero; however, given the high
value of the shear rate, the velocity distribution function rapidly
changes and adapts itself to the imposed shear rate, giving rise to
a sharp maximum of the reduced shear stress and normal stress
difference. Henceforth, as the temperature increases so does the
collision rate, so that the \textit{relative} strength of the shear
rate becomes smaller and smaller until the steady state is reached
at $s\approx 40$. The first stage lasts about one collision per
particle, has a \text{kinetic} nature, and is sensitive to the
initial preparation of the system. On the other hand, the subsequent
relaxation toward the steady state defines a much longer
\textit{hydrodynamic} stage that becomes more and more independent
of the initial state, provided that $T^0<T_\s$.  A similar
hydrodynamic regime exists for the class of initial states with
$T^0>T_\s$ (e.g., for $a\tau^0=0.1$). These comments also apply to
the two cases with $\alpha=0.5$.

   \begin{figure}
\includegraphics[width=1. \columnwidth]{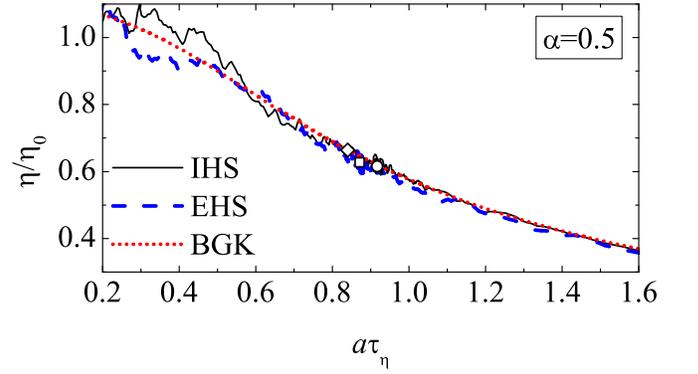}
 \caption{(Color online)
 (Transient) reduced shear viscosity $\eta^*(t)\equiv\eta(t)/\eta_0(t)$ versus the (transient) reduced
shear rate $a^*(t)\equiv a\tau_\visc(t)$ for IHS (solid lines) and EHS
(dashed lines) in the case $\alpha=0.5$.  The dotted lines are the
predictions obtained from the solution of a BGK model. The circle,
square, and diamond are the steady-state points for IHS, EHS, and
BGK, respectively. In each case, the curve to the left of the symbol
corresponds to $a\tau^0=0.1$, while the curve to the right of
the  symbol corresponds to $a\tau^0=4$.
\label{nonlinear_shear}}
 \end{figure}
In general, at a given value of
$\alpha$, the \textit{intrinsic} velocity distribution function in
the hydrodynamic regime depends on time only through its dependence on the shear rate nondimensionalized
with the (time-dependent) collision rate. More specifically,
\beq
f(\mathbf{V},t)=n \left[\frac{m}{2T(t)}\right]^{3/2}
f^*(\mathbf{C}(t);a^*(t)),
\label{V.1}
\eeq
where
\beq
\mathbf{C}(t)=\frac{\mathbf{V}}{\sqrt{2T(t)/m}}, \quad
a^*(t)=a\tau_\visc(t).
\label{V.2}
\eeq
The definition of the reduced shear rate $a^*$ by means of the
viscosity characteristic time $\tau_\visc$ instead of the
mean free time $\tau_{\text{mft}}$ or the characteristic time $\tau$
is an irrelevant matter of choice since these three quantities differ only by constant factors, namely
$\tau_{\text{mft}}/\tau_\visc\simeq 0.787$ and $\tau/\tau_\visc\simeq 0.888$. In the steady state, $a^*(t)\to
a^*_\s$ and $f^*(\mathbf{C};a^*)\to f^*(\mathbf{C};a_\s^*)=f_\s^*(\mathbf{C})$ [cf.\ Eq.\ \ref{III.12})].

During the
hydrodynamic relaxation  toward the steady state, it is insightful to define a
time-dependent shear viscosity $\eta(t)=-P_{xy}(t)/a$. As a
consequence of Eq. (\ref{V.1}), the ratio $\eta(t)/\eta_0(t)$, where
$\eta_0(t)=nT(t)\tau_\visc(t)$ is the Navier--Stokes
viscosity in the elastic limit [cf.\ Eq.\ (\ref{2.11.3})], depends
on time only through a nonlinear dependence on $a^*(t)$. Figure
\ref{nonlinear_shear} shows the reduced shear viscosity
$\eta^*\equiv \eta/\eta_0$ versus the reduced shear rate $a^*$ in
the case $\alpha=0.5$. The values plotted correspond to a temporal
window $2\lesssim s\lesssim 50$. The lower limit guarantees that the
system has practically lost memory of its initial state, while the
upper limit is long enough to guarantee that the steady state
(represented by an open symbol) has been reached. For each case
(IHS, EHS, or BGK), the steady-state point $(a_\s^*,\eta_\s^*)$ splits
the respective curve into two branches: the one to the left of the
point corresponds to $T<T^0$ (e.g., $a\tau^0=0.1$), while the
branch to the right corresponds to $T>T^0$ (e.g., $a\tau^0=4$).
The entire curve represents the \textit{nonlinear} shear viscosity
$\eta^*(a^*)$ for $\alpha=0.5$, and the steady-state point
$\eta_\s^*=\eta^*(a_\s^*)$ is just a particular (and singular) value \cite{SGD03}. The formal
extrapolation of $\eta^*(a^*)$ to zero shear rate gives the
(reduced)  Navier-Stokes shear viscosity
$\eta_{\text{NS}}^*=\lim_{a^*\to 0}\eta^*(a^*)$ at $\alpha=0.5$. The
expected values are \cite{SA04,BDKS98} $\eta_{\text{NS}}^*\simeq 1.3$ for
IHS and $\eta_{\text{NS}}^*\simeq 1.1$ for EHS and BGK. It is
worthwhile noting that, except for fluctuations in the simulation
data, the curves $\eta^*(a^*)$ for IHS, EHS, and BGK practically
coincide, at least in the interval $0.2<a^*<1.6$. Thus, the main
difference among the three approaches lies in the steady-state point where the
system ``decides'' to stop. For a more extensive discussion on the
rheological function $\eta^*(a^*)$ and the distinction between
$\eta_\s^*$ and $\eta_{\text{NS}}^*$ the reader is referred to Ref.\
\cite{SGD03}.

\begin{figure}
\includegraphics[width=1. \columnwidth]{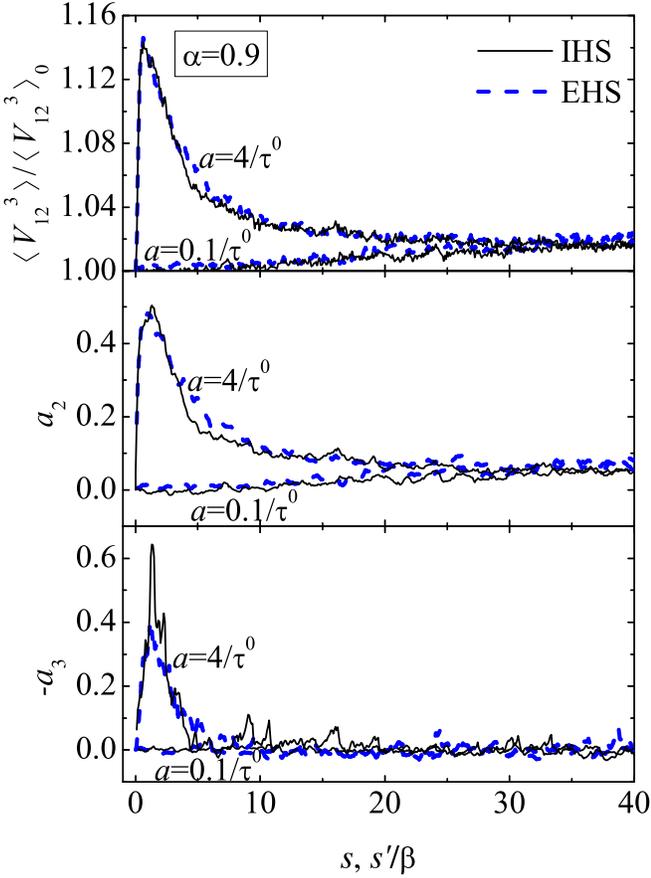}
 \caption{(Color online)
Evolution of  the ratio $\langle V_{12}^3\rangle/\langle
V_{12}^3\rangle_0$, the fourth cumulant $a_2$, and the sixth
cumulant $-a_3$ for IHS (solid lines) and EHS (dashed lines) in the
cases $\alpha=0.9$ with $a\tau^0=0.1$ and $a\tau^0=4$.
The data have been obtained starting from the initial distribution function  (\protect\ref{III.7}).
\label{hom_a2_09}}
 \end{figure}
  \begin{figure}
\includegraphics[width=1. \columnwidth]{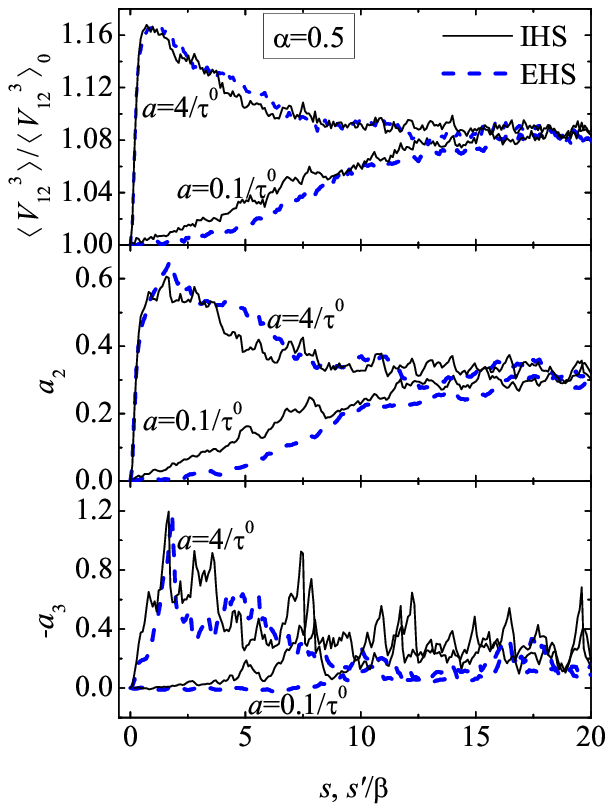}
 \caption{(Color online)
Evolution of  the ratio $\langle V_{12}^3\rangle/\langle
V_{12}^3\rangle_0$, the fourth cumulant $a_2$, and the sixth
cumulant $-a_3$ for IHS (solid lines) and EHS (dashed lines) in the
cases $\alpha=0.5$ with $a\tau^0=0.1$ and $a\tau^0=4$.
The data have been obtained starting from the initial distribution function  (\protect\ref{III.7}).
\label{hom_a2_05}}
 \end{figure}
   \begin{figure}
\includegraphics[width=1. \columnwidth]{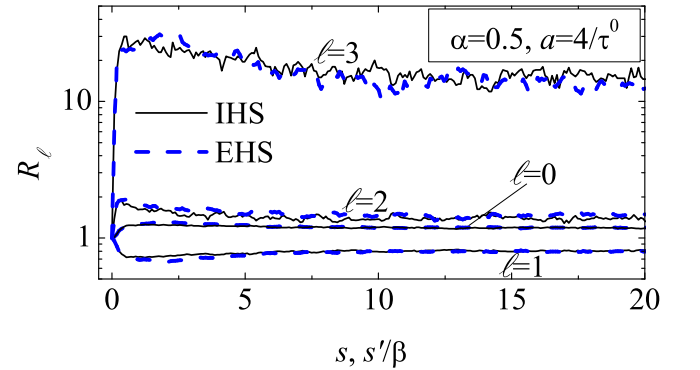}
 \caption{(Color online)
Evolution of  the ratios $R_0$, $R_1$, $R_2$, and $R_3$ [cf.\ Eq.\ (\protect\ref{new4})] for IHS (solid lines) and EHS (dashed lines) in the
case $\alpha=0.5$ with $a\tau^0=4$.
The data have been obtained starting from the initial distribution function  (\protect\ref{III.7}).
\label{vdf_evol}}
 \end{figure}
Although the most relevant information in the USF problem is
conveyed by the elements of the pressure tensor, they of course do
not exhaust the physical information one can extract from the
simulations. In Figs.\ \ref{hom_a2_09} and \ref{hom_a2_05} we
present the evolution of the ratio $\langle V_{12}^3\rangle/\langle
V_{12}^3\rangle_0$ (where $V_{12}$ is the relative speed between  a
pair of particles) and the fourth and sixth cumulants  [cf.\
Eq.\ (\ref{III.19})] for $\alpha=0.9$ and $\alpha=0.5$, respectively. The
average value $\langle V_{12}^3\rangle$ is physically interesting because it is
proportional to the true cooling rate $\zeta$ of the IHS gas. In
fact, we have checked in the simulations that the value of $\zeta$ obtained from Eq.\
(\ref{3.5}) agrees with the one computed directly from the
collisional energy loss, Eq.\ (\ref{3.4}). However, the second method is much noisier
than the first one since  it involves only the colliding  pairs,
whereas all the pairs contribute to $\langle V_{12}^3\rangle$ [see,
however, Eq.\ (\ref{IV.18}) and the comment above it].

As happened
with Figs.\ \ref{hom_hidro09} and \ref{hom_hidro05}, the simulation
data for each one of the three plotted quantities in the case
$a\tau^0=4$ present a respective maximum during the kinetic
stage of the evolution, representing the largest departure from the
Maxwellian.
It is noteworthy that the fluctuations of the quantities plotted in
Figs.\ \ref{hom_a2_09} and \ref{hom_a2_05} are much larger in the
IHS case  than in the EHS case. Otherwise, the temporal evolution
and the steady-state values are very similar in both systems.

We observe that $\langle V_{12}^3\rangle>\langle
V_{12}^3\rangle_0$, the relative difference increasing with the inelasticity. This
explains that the (internal) collisional cooling rate $\zeta$ of the IHS gas
is larger than the (external) frictional cooling rate $\zeta_0$ imposed on the EHS gas,
as said before in connection with Fig.\ \ref{hom_hidro05}. However,
even though the  imposed cooling rate for EHS is the local
equilibrium one, this system satisfactorily mimics the distortion
from local equilibrium, as measured by  $\langle
V_{12}^3\rangle/\langle V_{12}^3\rangle_0$, of the IHS system.

The cumulants basically probe the high-energy tail of the distribution, especially in the case of $a_3$.
The positive value of the fourth cumulant $a_2$, both for IHS and
EHS,  is a reflection of a strong high-energy overpopulation (with respect to the Maxwellian)  induced by
the shearing. This overpopulation effect is larger than and
essentially independent of the  one typically present in homogeneous
states of granular gases  \cite{vNE98,MS00}, which is absent in the
EHS gas. The steady-state value of the sixth cumulant $a_3$ is
practically zero for $\alpha=0.9$. On the other hand, for
$\alpha=0.5$ one has $-a_3>0$, what is again related to the
high-energy overpopulation.

A more direct information about the evolution of the velocity
distribution is provided by the ratios $R_\ell$ defined by Eq.\
(\ref{new4}). They are plotted in Fig.\ \ref{vdf_evol} for the case
$\alpha=0.5$ and $a\tau^0=4$. The maximum deviation from the
Maxwellian takes place during the kinetic stage ($s\lesssim 1$).
Thereafter, the curves smoothly relax toward their steady-state
values $R_0\simeq 1.18$, $R_1\simeq 0.81$, $R_2\simeq 1.39$, and $R_3\simeq
15$, the relaxation times being practically the same ($s\approx 10$) in the four cases.
During the hydrodynamic transient regime and in the steady state,
the population of particles moving with a speed larger than
three times the thermal speed $v_0(t)=\sqrt{2T(t)/m}$ is remarkably larger than the one
expected from a Maxwellian distribution. This overpopulation effect
induced by shearing
is also present in the interval $2v_0(t)<V<3v_0(t)$. In the hydrodynamic transient regime, between 90\% and
93\% of the particles move with a speed smaller than $2v_0(t)$ (in contrast to the equilibrium value of 95.4\%)
 and this is then the relevant region
for the low-degree moments. While the low-velocity region $V<v_0(t)$
is about 20\% overpopulated with respect to the Maxwellian, the
intermediate region $v_0(t)<V<2v_0(t)$ is underpopulated by about
the same amount. In fact, the fraction of particles moving with
$V<v_0(t)$ and $v_0(t)<V<2v_0(t)$ is about 51\% and 42\%,
respectively, in the steady state, while the corresponding
equilibrium values are 42.8\% and 52.6\%, respectively.
As Fig.\ \ref{vdf_evol} shows, all these features are successfully
captured by the EHS system, even at this rather high inelasticity.

The fact that $\langle V_{12}^3\rangle>\langle
V_{12}^3\rangle_0$
suggests that a better agreement between the dynamics of the IHS
gas and that of the EHS gas could be expected if the friction
constant of the latter were not chosen as $\gamma=\frac{1}{2}\zeta_0$ but as
$\gamma=\frac{1}{2}\zeta_0\langle V_{12}^3\rangle/\langle V_{12}^3\rangle_0$,
so that the cooling rate of the EHS gas would be exactly the same functional of $f$ as that of the IHS gas.
To test this expectation, we have carried out supplementary
simulations of the EHS system with this more refined friction
constant in the case $\alpha=0.5$, $a\tau^0=4$. The corresponding
curves are not included in Figs.\ \ref{hom_coll}, \ref{hom_coll2},
\ref{hom_hidro05}, \ref{nonlinear_shear}, \ref{hom_a2_05}, and \ref{vdf_evol} for
the sake of clarity. The results show that, in general, the quantities associated with the low-order moments
(temperature and pressure tensor) are indeed closer to the IHS values
than with the simpler choice $\gamma=\frac{1}{2}\zeta_0$. As a
matter of fact, the EHS temperature is now slightly smaller, instead of slightly larger,
than the IHS temperature. However, in the cases of $\langle
V_{12}^3\rangle/\langle
V_{12}^3\rangle_0$, $a_2$, and $a_3$, the results obtained in the
EHS simulations with $\gamma=\frac{1}{2}\zeta_0\langle V_{12}^3\rangle/\langle
V_{12}^3\rangle_0$ turn out to be not necessarily better than those
obtained with $\gamma=\frac{1}{2}\zeta_0$.

\subsection{Steady state\label{sec5C}}
\begin{figure}
\includegraphics[width=1. \columnwidth]{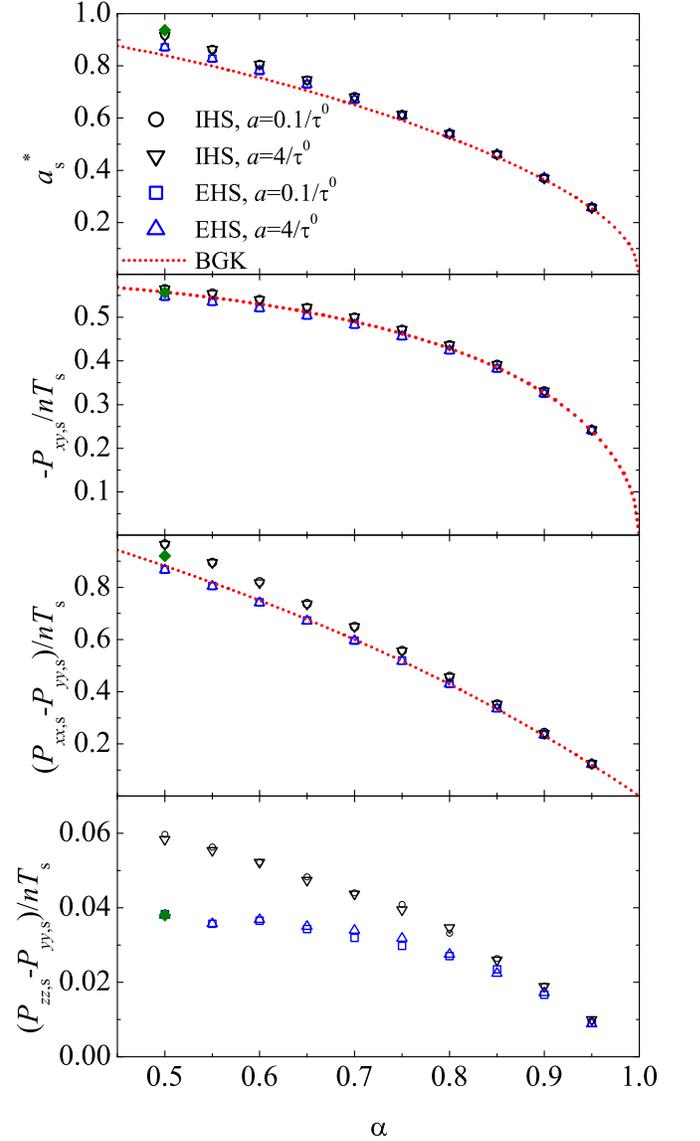}
 \caption{(Color online)
Steady-state values of the reduced shear rate
$a^*=a\tau_\visc$, the reduced shear stress $-P_{xy}/nT$,
and the reduced normal stress differences $(P_{xx}-P_{yy})/nT$ and
$(P_{zz}-P_{yy})/nT$ as functions of the coefficient of restitution
$\alpha$. The open symbols are simulation results for IHS and EHS and two
values of the shear rate, while the dotted lines correspond to the
solution of a BGK model. Note that in the latter model
$P_{zz,\s}=P_{yy,\s}$.
The filled diamonds represent simulation data of EHS with a friction
constant $\gamma=\frac{1}{2}\zeta_0\langle V_{12}^3\rangle/\langle
V_{12}^3\rangle_0$ in the case $\alpha=0.5$, $a\tau^0=4$.
\label{hidro_ss}}
 \end{figure}
Once we have analyzed the transient period toward the steady state
for the representative cases $\alpha=0.5$  and $\alpha=0.9$, let us
report on the most relevant steady-state properties for all the
values $\alpha=0.5$--$0.95$ we have considered. The quantities
associated with the second-degree velocity moments, namely the
reduced shear rate $a_\s^*=a\tau_\visc(T_\s)\propto 1/\sqrt{T_\s}$, the reduced shear
stress $-P_{xy,\s}/nT_\s$, and the reduced normal stress differences
$(P_{xx,\s}-P_{yy,\s})/nT_\s$ and $(P_{zz,\s}-P_{yy,\s})/nT_\s$, are plotted in Fig.\
\ref{hidro_ss}. The overlapping of the data obtained from
simulations with the two different values of the shear rate
($a\tau^0=0.1$  and $a\tau^0=4$) confirms that the steady
state has actually been reached and that the intrinsic velocity
distribution function $f^*_\s(\mathbf{C})$ [cf.\ Eq.\ (\ref{III.12})]
depends only on $\alpha$ and not on the initial preparation of the
system. For $\alpha\gtrsim 0.7$ there exists a good agreement between
the EHS and IHS results for the quantities $a_\s^*$ and
$-P_{xy,\s}/nT_\s$, which are the most relevant properties in the USF
problem. For larger inelasticity, however, the steady-state
temperature $T_\s$ is larger for the EHS gas than for the IHS gas,
and so $a_\s^*$ is smaller in the former case than in the latter.
This implies that the departure from isotropy is slightly smaller in
the EHS gas than in the IHS gas, and so is the shear stress
$-P_{xy,\s}/nT_\s$. As for the normal stress differences
$(P_{xx,\s}-P_{yy,\s})/nT_\s$ and, especially, $(P_{zz,\s}-P_{yy,\s})/nT_\s$, they
start to differ in both systems for $\alpha\lesssim 0.85$. It is
worthwhile noting  that the BGK kinetic model, which has a simple explicit
solution in the steady state \cite{SA04,SGD03}, does a very good job at
predicting the transport properties, especially in the case of the
EHS system. An exception is the normal stress difference
$(P_{zz,\s}-P_{yy,\s})/nT_\s$, which vanishes in the BGK model but
takes on (small) positive values in the simulations. The good
agreement between simulation data for the elements of the pressure
tensor  and the BGK predictions was already noted in Ref.\
\cite{BRM97}, although there the kinetic model was slightly
different from the one considered here.

Figure \ref{hidro_ss} also includes results obtained from  control
simulations carried out in the case $\alpha=0.5$ on EHS
but with a refined friction constant $\gamma=\frac{1}{2}\zeta_0\langle V_{12}^3\rangle/\langle
V_{12}^3\rangle_0$ instead of the simple one
$\gamma=\frac{1}{2}\zeta_0$. We observe that the agreement with the IHS data improves for those quantities that
were already reasonably well described by the simple EHS system,
namely $T_\s$, $P_{xy,\s}$, and $P_{xx,\s}-P_{yy,\s}$. On the other
hand, the delicate normal stress difference $P_{zz,\s}-P_{yy,\s}$,
which otherwise is quite small, is still about 40\% smaller in both EHS
systems than in the IHS system. This indicates that whenever the
discrepancies between the IHS and EHS results are relatively
important, they are hardly affected by a more sophisticated choice
of the friction constant $\gamma$.

  \begin{figure}
\includegraphics[width=1. \columnwidth]{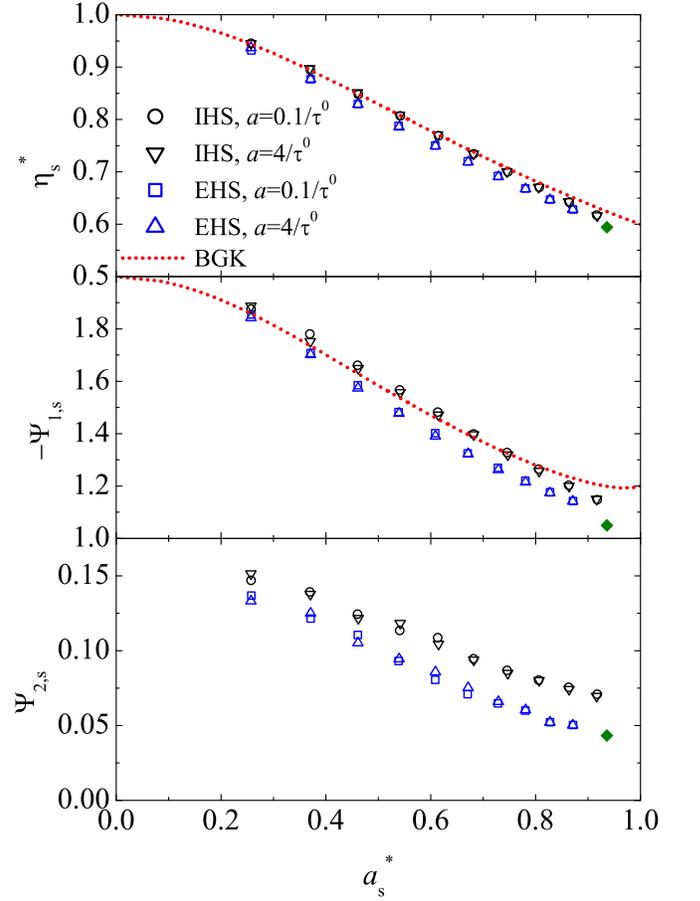}
 \caption{(Color online)
Steady-state values of the reduced shear viscosity
$\eta^*=\eta/\eta_0$ and the viscometric functions
$-\Psi_1=(P_{xx}-P_{yy})/nT{a^*}^2$ and
$\Psi_2=(P_{zz}-P_{yy})/nT{a^*}^2$  as functions of the reduced
shear rate $a_\s^*$. The open symbols are simulation results for IHS and
EHS and two values of the shear rate, while the dotted lines
correspond to the solution of a BGK model. Note that in the latter
model $\Psi_{2,\s}=0$. The filled diamonds represent simulation data of EHS with a friction
constant $\gamma=\frac{1}{2}\zeta_0\langle V_{12}^3\rangle/\langle
V_{12}^3\rangle_0$ in the case $\alpha=0.5$, $a\tau^0=4$.
\label{reol_ss}}
 \end{figure}
 {}From a rheological point of view, it is worthwhile introducing the
nonlinear shear viscosity $\eta^*=-(P_{xy}/a)/\eta_0=-(P_{xy}/nT)/a^*$ and the
viscometric functions $\Psi_1=(P_{yy}-P_{xx})/nT{a^*}^2$ and
$\Psi_2=(P_{zz}-P_{yy})/nT{a^*}^2$. In the hydrodynamic transient
regime,  they are functions of $a^*$ (for a given value of $\alpha$), as
was  illustrated in Fig.\ \ref{nonlinear_shear} in the case of
$\eta^*$ at $\alpha=0.5$. In the steady state ($a^*\to a^*_\s$) these  quantities become functions
of $\alpha$ only. Equivalently, by eliminating $\alpha$ in favor of
the  steady-state reduced shear rate $a^*_\s$, those rheological
quantities can be seen as functions of $a^*_\s$. This is the representation shown in
Fig.\ \ref{reol_ss}. We observe that the curve $\eta_\s^*(a_\s^*)$ for
EHS is very close to the one for IHS, even though the EHS points are
shifted with respect to the IHS points corresponding to the same
value of $\alpha$, the shift increasing as $\alpha$ decreases. On the other hand, the viscometric effects are
more pronounced in the IHS case than in the EHS case. Paradoxically,
 the BGK curves for
$\eta^*_\s$ and $\Psi_{1,\s}$ are generally closer to IHS than to EHS in the representation of Fig.\ \ref{reol_ss}.
It is also noteworthy that the isolated points corresponding to EHS
with $\gamma=\frac{1}{2}\zeta_0\langle V_{12}^3\rangle/\langle
V_{12}^3\rangle_0$ seems to be consistent with the curve obtained by
joining the other EHS points.

  \begin{figure}
\includegraphics[width=1. \columnwidth]{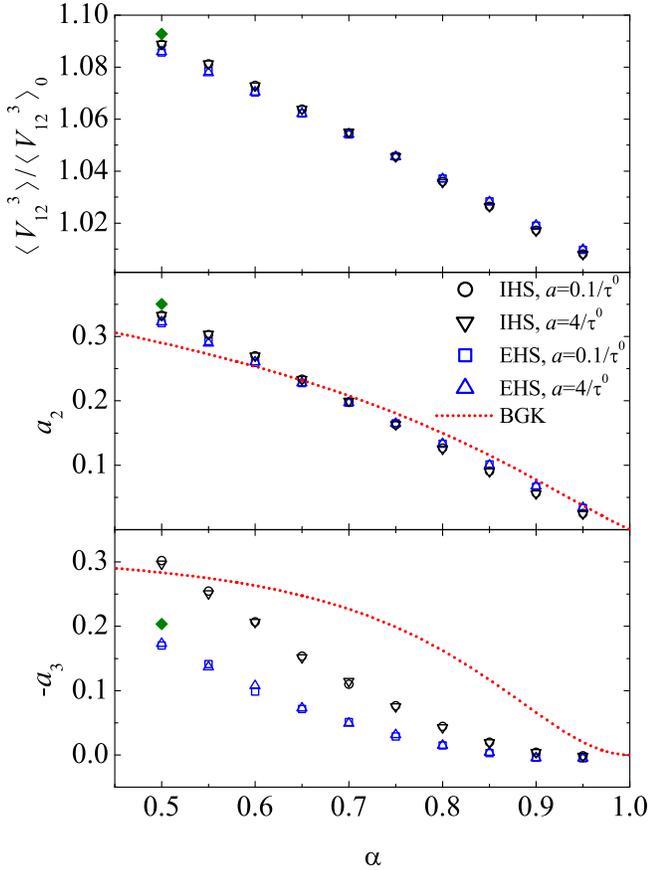}
 \caption{(Color online)
Steady-state values of the the ratio $\langle
V_{12}^3\rangle/\langle V_{12}^3\rangle_0$, the fourth cumulant
$a_2$, and the sixth cumulant $-a_3$  as functions of the
coefficient of restitution $\alpha$. The open symbols are simulation
results for IHS and EHS and two values of the shear rate, while the
dotted lines in the middle and bottom panels correspond to the
solution of a BGK model. The filled diamonds represent simulation data of EHS with a friction
constant $\gamma=\frac{1}{2}\zeta_0\langle V_{12}^3\rangle/\langle
V_{12}^3\rangle_0$ in the case $\alpha=0.5$, $a\tau^0=4$.
\label{a2_ss}}
 \end{figure}
The $\alpha$-dependence of $\langle V_{12}^3\rangle/\langle
V_{12}^3\rangle_0$ and the cumulants $a_2$ and $a_3$ in the steady state is displayed in
Fig.\ \ref{a2_ss}. We recall that in the IHS system the ratio
$\langle V_{12}^3\rangle/\langle V_{12}^3\rangle_0$ coincides with
the ratio $\zeta/\zeta_0$ between the true cooling rate and its
local equilibrium estimate. It is observed that the local equilibrium
approximation underestimates the cooling rate  by a few percent, essentially due to the distortion induced by the
shearing. This distortion is well captured by the EHS system,
despite the fact that its cooling rate is, by construction, given by
$\zeta_0$. As said above, the positive values of $a_2$ and, especially, $-a_3$
are indicators of an  overpopulation  effect (with respect to the
Maxwellian) of the high-velocity tail of the distribution, this effect
being stronger than and practically independent of the typical high-velocity
overpopulation of granular gases in uniform and isotropic states
\cite{vNE98,MS00}. As a matter of fact, the distribution function of
the (frictional) EHS gas in the homogeneous cooling state, as well
as in the state heated by a white-noise forcing, is exactly a
Gaussian. Therefore, the overpopulation measured by $a_2$ and $-a_3$
is basically a shearing effect. The fourth cumulant is well
accounted for by the EHS gas, but  the magnitude of the sixth
cumulant is larger for IHS than for EHS. As happens with the normal
stress difference $P_{zz}-P_{yy}$ (see the bottom panel of Fig.\ \ref{hidro_ss}), the cumulant $a_3$ is a very
sensitive quantity that probes subtle details of the IHS velocity distribution
function not sufficiently well captured by the EHS system for
$\alpha\lesssim 0.85$.

It is worthwhile noting that
$\langle V_{12}^3\rangle/\langle V_{12}^3\rangle_0$ and $a_2$ are
slightly larger in the EHS case than in the IHS case for
$\alpha\gtrsim 0.7$, while the opposite happens for $\alpha\lesssim
0.7$. This is reminiscent of the situation in the homogeneous
cooling state and in the  white-noise heated state \cite{vNE98},
where $a_2^{\text{IHS}}\leq
a_2^{\text{EHS}}=0$ for $\alpha\geq \sqrt{2}/2\simeq 0.71$,
while $ a_2^{\text{IHS}}\geq
a_2^{\text{EHS}}=0$ for $\alpha\leq \sqrt{2}/2$. In fact, we
have observed that the difference $ a_2^{\text{IHS}}-
 a_2^{\text{EHS}}$ in the USF is quite close to the
difference in the white-noise heated state. Figure \ref{a2_ss} also
shows a strong correlation between the values of $\langle
V_{12}^3\rangle/\langle V_{12}^3\rangle_0$ and those of $a_2$. More
precisely, we have empirically checked that $\langle V_{12}^3\rangle/\langle
V_{12}^3\rangle_0\simeq 1+0.27 a_2$  both for IHS and EHS in USF, in
contrast to $\langle V_{12}^3\rangle/\langle V_{12}^3\rangle_0\simeq
1+\frac{3}{16} a_2$ in isotropic states \cite{vNE98}.

In Fig.\
\ref{a2_ss} we have included the BGK predictions for the cumulants
$a_2$ and $a_3$. Not being associated with  conventional velocity
moments, the evaluation of $\langle V_{12}^3\rangle$ [cf.\ Eq.\
(\ref{3.5bis})] from the BGK distribution function requires heavy
numerical work and so it is not included in Fig.\ \ref{a2_ss}. We
observe that the increase of $a_2$ with increasing inelasticity is
well captured by the BGK model. On the other hand, the
$\alpha$-dependence of $-a_3$ is described by the BGK model at a
qualitative level only, predicting in general too high values.

Regarding the control simulations for EHS with $\alpha=0.5$ and $\gamma=\frac{1}{2}\zeta_0\langle V_{12}^3\rangle/\langle
V_{12}^3\rangle_0$, we observe that the values of $\langle V_{12}^3\rangle/\langle
V_{12}^3\rangle_0$ and $a_2$ are not actually improved, while the improvement
on $a_3$ is very small.

\begin{figure*}
\includegraphics[width=2. \columnwidth]{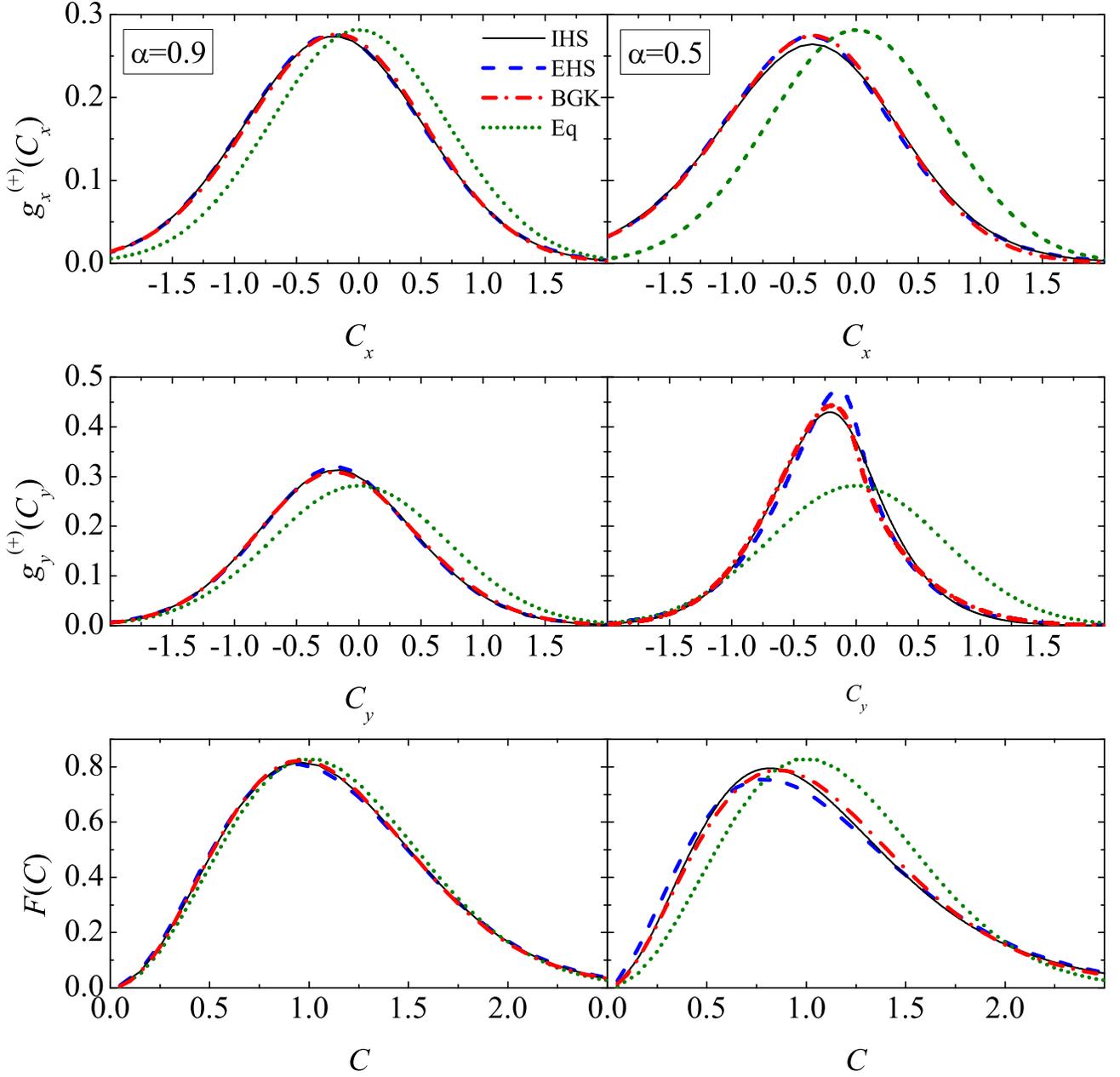}
 \caption{(Color online)
 Linear plots of the marginal velocity distribution functions $g_x^{(+)}(C_x)$,
 $g_y^{(+)}(C_y)$, and $F(C)$ for $\alpha=0.9$ (left panels) and $\alpha=0.5$ (right
 panels). The solid and dashed lines represent simulation results
 for IHS and EHS, respectively, the dash-dot lines are the BGK
 predictions, and the dotted lines are the (local) equilibrium
 distributions.
\label{vdf0}}
 \end{figure*}
\begin{figure*}
\includegraphics[width=2. \columnwidth]{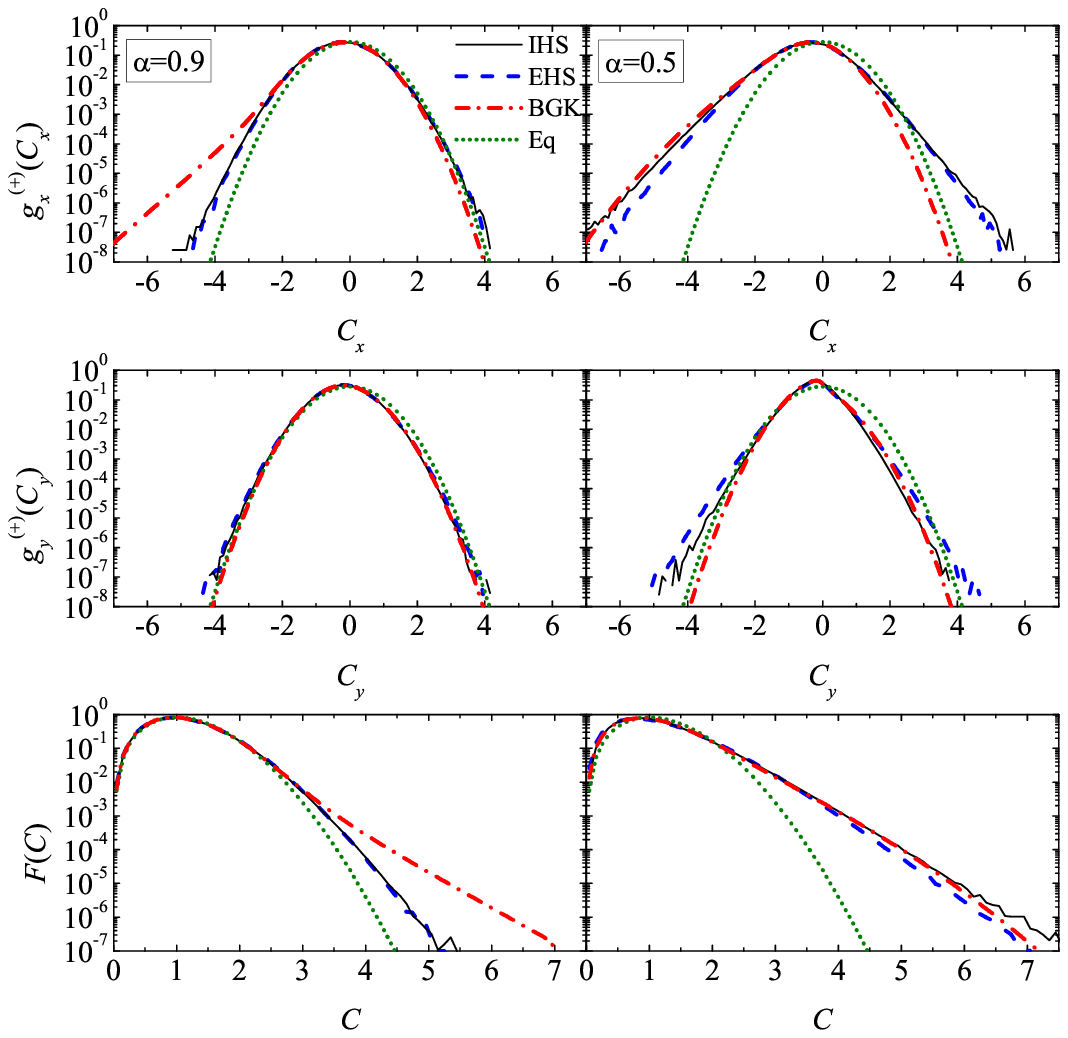}
 \caption{(Color online)
 Logarithmic plots of the marginal velocity distribution functions $g_x^{(+)}(C_x)$,
 $g_y^{(+)}(C_y)$, and $F(C)$ for $\alpha=0.9$ (left panels) and $\alpha=0.5$ (right
 panels). The solid and dashed lines represent simulation results
 for IHS and EHS, respectively, the dash-dot lines are the BGK
 predictions, and the dotted lines are the (local) equilibrium
 distributions.
\label{vdf}}
 \end{figure*}
The quantities plotted in Figs.\ \ref{hidro_ss}--\ref{a2_ss} provide
useful (indirect) information about the velocity distribution
function of the steady-state USF. However, they essentially probe
the domain of low and moderate velocities (say $V\lesssim 2
\sqrt{2T/m}$), except perhaps in the cases of $a_2$ and, especially,
$a_3$, which are more sensitive to the high-velocity tail. In order
to analyze more directly the shape of the velocity distribution, its
anisotropy, and the high-velocity tail, we have measured in the
simulations the steady-state marginal distributions defined by Eqs.\
(\ref{III.15})--(\ref{III.14}). As representative examples, Figs.\
\ref{vdf0} (linear scale)  and \ref{vdf} (logarithmic scale)  show
$g_x^{(+)}(C_x)$,
 $g_y^{(+)}(C_y)$, and $F(C)$ for $\alpha=0.9$  and $\alpha=0.5$. We
 have checked that the symmetry properties (\ref{III.17}) are fulfilled and that the
 curves obtained from the two values of the shear rate
 ($a\tau^0=4$ and $a\tau^0=0.1$) practically coincide. In fact, to
 improve the statistics, the simulation data represented in Fig.\
 \ref{vdf} have been averaged over both shear rates and, in
 addition, the symmetry properties (\ref{III.17}) have been exploited
 to make
 \beq
 \begin{array}{l}
 g_x^{(+)}(C_x)\to
 \frac{1}{2}\left[g_x^{(+)}(C_x)+g_x^{(-)}(-C_x)\right],\\
 g_y^{(+)}(C_y)\to
 \frac{1}{2}\left[g_y^{(+)}(C_y)+g_y^{(-)}(-C_y)\right].
 \end{array}
 \label{5.1}
 \eeq
Two anisotropic features of the USF state are quite apparent. First, the
functions  $g_x^{(+)}(C_x)$ and $g_y^{(+)}(C_y)$ are clearly
asymmetric, namely $g_x^{(+)}(|C_x|)<g_x^{(+)}(-|C_x|)$ and
$g_y^{(+)}(|C_y|)<g_y^{(+)}(-|C_y|)$. This is a physical effect
induced by the shearing, in consistency with $P_{xy}\propto\langle C_x
C_y\rangle<0$. The second feature is the non-Newtonian property
$g_x^{(+)}\neq g_y^{(+)}$. More specifically, the marginal
distribution $g_x^{(+)}$ is broader than $g_y^{(+)}$, in agreement
with the fact that $P_{xx}-P_{yy}\propto\langle C_x^2\rangle-\langle
C_y^2\rangle>0$. These two effects are obviously more pronounced for
$\alpha=0.5$ than for $\alpha=0.9$.

Figure \ref{vdf0} shows that  an excellent
 agreement  between the IHS and EHS distributions for an inelasticity
 $\alpha=0.9$ exits in the region of
 low and intermediate velocities, in consistency with  the results displayed in Figs.\
 \ref{hidro_ss}--\ref{a2_ss}. For $\alpha=0.5$, it is
 observed that the EHS distributions $g_x^{(+)}(C_x)$,
 $g_y^{(+)}(C_y)$, and $F(C)$ are more populated than the IHS ones
 in the regions $-1\lesssim C_x\lesssim 0$, $-0.3\lesssim C_y\lesssim
 0$, and $C\lesssim 0.5$, respectively.
It is also interesting to note that, in the region of thermal
velocities, the orientation-averaged distribution function $F(C)$ is
much less distorted with respect to the Maxwellian than the marginal
distributions $g_x^{(+)}(C_x)$ and $g_y^{(+)}(C_y)$, especially in
the case $\alpha=0.9$. Another important finding is that the BGK
kinetic model, not only  captures well ``global'' or average
properties, such as the hydrodynamic quantities (cf.\ Figs.\
\ref{hidro_ss}--\ref{a2_ss}), but also the ``local'' details of the
velocity distribution function.

The logarithmic scale employed in Fig.\ \ref{vdf} has been chosen to
reveal the high-velocity tails of the distributions. We observe an
overpopulation of both tails of $g_x^{(+)}$, especially at
$\alpha=0.5$. In the case of $g_y^{(+)}$, however, the
overpopulation seems to affect the tail $C_y<0$ only. The
high-energy overpopulation is clearly apparent in the distribution
of the magnitude of  the velocity, $F(C)$. In a recent paper,
Bobylev et al.\ \cite{BGP04} have proven that $\ln F(C)\sim -C^\mu$,
with $\mu\geq 1$, for asymptotically large velocities in the USF.
Although it is not clear whether our simulation data have reached
the high-velocity regime where the asymptotic law $\ln F(C)\sim
-C^\mu$ dominates, the bottom panels of Fig.\ \ref{vdf} seem to be
consistent with this law with $\mu\simeq 1$.

The comments in the preceding paragraph apply equally to IHS and
EHS. As a matter of fact, the simulation data for both systems in
the case $\alpha=0.9$ are hardly distinguishable for that value of
$\alpha$. In the case $\alpha=0.5$, however, the high-velocity
values of $g_x^{(+)}(C_x)$ and $F(C)$ are larger for IHS than for
EHS, while the opposite happens for the high-velocity values of
$g_y^{(+)}(C_y)$. As for the BGK distribution function, it cannot be
expected to be accurate beyond the domain of thermal velocities.
This is confirmed by Fig.\ \ref{vdf}, where we can observe that at
$\alpha=0.9$ the BGK model strongly exaggerates the high-velocity
overpopulation effects of $g_x^{(+)}(C_x)$ and $F(C)$. This produces
too large a value of the sixth cumulant $-a_3$, as observed in Fig.\
\ref{a2_ss}. On the other hand, at $\alpha=0.5$ the BGK value of
$-a_3$ agrees by accident with that of the IHS system and this
explains why in that case the overpopulated tail predicted by the
BGK model agrees well with that of the IHS system. Nevertheless, the
tail of $g_x^{(+)}(C_x)$ for
 $C_x>0$ as well as that of $g_y^{(+)}(C_y)$ for $C_y<0$ are
 strongly underestimated by the BGK model at $\alpha=0.5$.

 For the sake of clarity, we have not included in the right panels
 of Figs.\ \ref{vdf0} and \ref{vdf} the curves corresponding to our simulations of
 the EHS system with $\alpha=0.5$ and a friction constant $\gamma=\frac{1}{2}\zeta_0\langle V_{12}^3\rangle/\langle
V_{12}^3\rangle_0$. In any case, the results  are very close to
those corresponding to the friction constant
$\gamma=\frac{1}{2}\zeta_0$. This shows that the main quantitative
differences between highly dissipative IHS and EHS systems, in
particular the high-velocity tails, are intrinsic to the different
dynamics of both systems and so they are not avoided by a
fine-tuning of the drag force acting on the elastic particles.

\section{Conclusions\label{sec6}}

In this paper we have dealt with the main nonequilibrium properties
of two classes of dissipative gases subject to the so-called simple or uniform
shear flow (USF). In the first class, the system is made
of inelastic hard spheres (IHS) with a constant coefficient of
restitution $\alpha<1$. The inelasticity of collisions provides an
internal energy sink, characterized by a cooling rate $\zeta$. In
the second class, the particles are elastic hard spheres (EHS) under
the action of a drag force, so that the energy sink is now external
and characterized by a friction coefficient $\gamma$. At a given
value of $\alpha$, some of the basic properties of the Boltzmann
equation for the IHS gas are formally similar to those for the
(frictional) EHS gas \cite{SA04} if (i) the friction coefficient is
chosen as a function of the local density and temperature,
$\gamma=\frac{1}{2}\zeta_0\propto nT^{1/2}(1-\alpha^2)$, and (ii)
the collision rate of the EHS gas is
$\beta(\alpha)=\frac{1}{2}(1+\alpha)$ times the collision rate of
the IHS gas under the same conditions. The boundary conditions of
the USF state provides an energy source (viscous heating) that
competes with the energy dissipation (either collisional or
frictional), until a steady state is eventually reached, resulting
from the balance between both effects.

The two main points we have intended to address in this paper are
the following ones. On the one hand, we wanted to perform an
extensive study of the physical properties of  dissipative gases
under USF,
 for both the transient and the steady states.
While the USF state has been widely considered in the literature of
granular fluids, some of its relevant aspects (hydrodynamic transient stage,
cumulants, high-velocity tail, \ldots) have received little
attention. As a second point, taking the USF as a paradigmatic and
conceptually simple nonequilibrium state, we were interested in
elucidating to what extent the EHS gas mimics the physical
properties of the genuine IHS gas. To meet these goals, we have
carried out computer simulations by the DSMC method on both classes
of systems for ten values of the coefficient of restitution
($\alpha=0.5$--$0.95$ with a step $\Delta\alpha=0.05$) and two
values of the shear rate ($a=4/\tau^0$ and $a=0.1/\tau^0$).
Below we summarize the main conclusions derived from our study.

The duration of the transient period, when measured by the number of
collisions per particle ($s$), is hardly dependent on the imposed shear
rate $a$ or on the initial state, but strongly depends on the
coefficient of restitution $\alpha$. The larger the dissipation, the smaller
the number of collisions needed to reach the steady state. For
instance, $s\simeq 40$ at $\alpha=0.9$, while $s\simeq 20$ at
$\alpha=0.5$. Nevertheless, when the duration is measured by an
external clock (for instance, in units of the initial mean free
time), it becomes weakly dependent on $\alpha$ but strongly
dependent on $a$: the smaller the imposed shear rate, the longer the
transient period.

The evolution toward the steady state proceeds in two stages. The
first (kinetic) stage depends heavily on the initial preparation of
the system and lasts a few collisions per particle. This is followed
by a much slower hydrodynamic regime that becomes less and less
dependent on the initial state. Once conveniently scaled with the
thermal speed $v_0(t)=\sqrt{2T(t)/m}$, the velocity distribution
function in the hydrodynamic regime depends on time through the
reduced  velocity $\mathbf{C}(t)=\mathbf{V}/v_0(t)$ and the reduced
shear rate $a^*(t)\propto a/\sqrt{T(t)}$ only \cite{SGD03}. In particular, at a
given value of $\alpha$, the (reduced) nonlinear shear viscosity $\eta^*(a^*)$
moves on a certain rheological curve,  the steady-state value
$\eta_\s^*=\eta^*(a^*_\s)$ representing just one point. This point
splits the curve $\eta^*(a^*)$ into two branches. The branch
$a^*\leq a^*_\s$ is accessible from initial states such that the
dissipative cooling dominates over the viscous heating, so that
$T(t)$ decreases and $a^*(t)$ increases during the transient period.
Conversely, the branch
$a^*\geq a^*_\s$ corresponds to initial states where
the viscous heating dominates, so that
$T(t)$ increases and $a^*(t)$ decreases, until the steady state is reached.

The actual number of collisions per particle $s(t)$ is slightly smaller
than the local equilibrium estimate $s_0(t)$. This indicates that
$\langle V\rangle<\langle V\rangle_0$, whereas $\langle V^2\rangle=\langle
V^2\rangle_0=3T(t)/m$ by definition. The inequality $\langle V\rangle<\langle
V\rangle_0$ is consistent with an  underpopulation (with
respect to the local equilibrium distribution) for moderate velocities that must be
compensated by a high-velocity overpopulation. This qualitative
reasoning is confirmed by the inequalities $\langle V_{12}^3\rangle>\langle
V_{12}^3\rangle_0$,  $\langle V^4\rangle>\langle
V^4\rangle_0$, and $\langle V^6\rangle>\langle
V^6\rangle_0$ observed in the simulations. In addition to these effects, the shearing
motion induces a strong anisotropy in the normal stresses, namely
$P_{xx}(t)>nT(t)>P_{zz}(t)\gtrsim P_{yy}(t)$. In other words, a
breakdown of the energy equipartition occurs, whereby the
``temperature'' associated with the degree of freedom parallel to
the fluid motion is significantly larger than that associated with the
other two degrees of freedom, as already observed in previous studies \cite{CG86,WB86,JR88,C89,HS92,LB94,SGN96,BRM97,MGSB99,F00,BGS02,AL03}.

We have paid special attention to the properties of the steady
state, which is intrinsically independent of the imposed shear rate
and of the initial state. As expected, the distortion from the local
equilibrium state (as measured by the shear stress, the normal
stress differences, the cumulants, \ldots) increases with the
dissipation. This distortion is made quite apparent by the shapes of the steady-state (marginal) velocity
distributions defined by Eqs.\ (\ref{III.15})--(\ref{III.14}). In
particular, the high-velocity tails of $g_x^{(+)}(C_x)$ and $F(C)$
seem to be consistent with an exponential overpopulation.

All the above comments apply equally well to the IHS and EHS gases.
Therefore,  the main nonequilibrium
and transport properties of a true IHS gas in the USF state are
satisfactorily mimicked by an ``equivalent'' EHS gas. If one focus on the basic properties (say the
steady-state reduced shear rate $a_\s^*$, shear stress
$P_{xy,\s}/nT_\s$, or  second cumulant $a_2$) it is almost
impossible to distinguish the EHS values from the IHS ones if
$\alpha\gtrsim 0.7$, the differences still being relatively small if $\alpha\lesssim 0.7$.
We have observed that $\langle V_{12}^3\rangle/\langle
V_{12}^3\rangle_0$ and $a_2$ are in the EHS system slightly larger than in
the IHS system if $\alpha\gtrsim 0.7$, while the opposite happens if
$\alpha\lesssim 0.7$; this is analogous to what happens in
 the homogeneous cooling state and in the
white-noise heated state, in which cases the EHS distribution is exactly a
Maxwellian.
It is interesting to note that, even at $\alpha=0.5$, the full hydrodynamic curve
$\eta^*(a^*)$ coincides for both systems, except that the location
of the steady-state point $\eta_\s^*=\eta^*(a^*_\s)$ slightly
changes.
More delicate quantities (such as the
normal stress differences or the sixth cumulant) keep being
practically the same in both systems if $\alpha\gtrsim 0.85$. Even
at the level of the velocity distribution function itself, the EHS
and IHS curves practically overlap (at least for the domain of
velocities $C\lesssim 6$ accessible to our computer simulations) at
a coefficient of restitution as realistic as $\alpha=0.9$. At
$\alpha=0.5$, however, the distribution functions $g_x^{(+)}(C_x)$ and $F(C)$ of the IHS
system exhibit a visibly larger high-velocity overpopulation than
those of the EHS system.

As said above, in the EHS systems we have chosen the friction coefficient as
$\gamma=\frac{1}{2}\zeta_0\propto nT^{1/2}(1-\alpha^2)$, so that it is a functional of the
distribution function only through the local density and temperature and,
moreover, its dependence on $\alpha$ is explicit. Given that the
true cooling rate is slightly larger than the local equilibrium
estimate, i.e., $\zeta/\zeta_0=\langle V_{12}^3\rangle/\langle
V_{12}^3\rangle_0>1$, the imitation of the inelastic cooling rate by
the EHS is not perfect. Therefore, one might reasonably expect that the
discrepancies between the EHS and IHS results would diminish if the
friction coefficient were taken as $\gamma=\frac{1}{2}\zeta_0\langle V_{12}^3\rangle/\langle
V_{12}^3\rangle_0$. To test this expectation, we have performed
complementary simulations of the EHS system with this more refined
value of $\gamma$ in the case of highest dissipation, i.e., $\alpha=0.5$. The results show that,
whenever the former agreement between EHS and IHS was fair, the new
agreement is generally even better. However, those quantities
(such as the normal stress difference $P_{zz}- P_{yy}$ and the sixth
cumulant $a_3$) that turned out to be especially sensitive to the dissipation mechanism
(collisional inelasticity versus external friction) are practically
unaffected by the new choice of $\gamma$. Moreover, the
high-velocity tails of both versions of the EHS gas are practically
the same, being both smaller than the IHS tail.
Therefore, the (subtle) discrepancies between the IHS and EHS systems
in cases of high dissipation (say $\alpha\lesssim 0.7$) seem to be intrinsic to their distinct
dynamics. Taking this into account, there is no practical reason to
propose for the drag force acting on the EHS a friction coefficient
different from the local equilibrium value
$\gamma=\frac{1}{2}\zeta_0$. Concerning the collision rate
coefficient $\beta(\alpha)$, the choice
$\beta(\alpha)=\frac{1}{2}(1+\alpha)$ is recommended by criteria of
simplicity and consistency with the cases of mixtures and dense
gases \cite{SA04}. Moreover, we have checked (not shown in this paper) that an alternative
choice, namely $\beta(\alpha)=\frac{1}{6}(1+\alpha)(2+\alpha)$,  although reproducing well the Navier--Stokes shear
viscosity \cite{SA04}, provides results for the nonlinear shear viscosity in worse agreement with the
IHS ones than those reported here with
$\beta(\alpha)=\frac{1}{2}(1+\alpha)$.

The (approximate) equivalence
IHS$\leftrightarrow$EHS can be used to transfer to granular gases part of the expertise
accumulated for a long time on the kinetic theory of elastic
particles. In particular, the celebrated Bhatnagar--Gross--Krook
(BGK)
kinetic model of the Boltzmann equation can be readily
extended to granular gases \cite{BDS99,SA04}. In this paper we have
compared the solution of the BGK model for USF
\cite{SA04,SGD03,BRM97} with the simulation data. While it is generally believed that
the BGK model  would be accurate only for states near equilibrium and/or in the quasi-elastic limit, our results
show that, despite its simplicity, the model succeeds in capturing
quantitatively the evolution and steady-state values of the main
transport properties (temperature, shear stress, and normal stress
difference $P_{xx}-P_{yy}$) and even of the fourth cumulant $a_2$.
However, the small nonzero difference $P_{zz}-P_{yy}$ is not
accounted for by the BGK model and the sixth cumulant $a_3$ agrees
with the simulation data at a qualitative level only. All of this is
consistent with the observation that the BGK velocity distribution
function is reliable in the thermal region ($C\lesssim 2$), but not
in the high-velocity domain.

In summary, we conclude that the solutions of the Boltzmann
equations
for IHS are very similar to those for EHS (in the latter case with a smaller collision rate and under the action of  an adequate drag
force). Thus the temporal evolution toward the steady state and the
properties of the latter are mainly governed by the common feature of
energy dissipation, without any significant influence of the
detailed mechanism behind it. Only for very high dissipation  (say $\alpha\lesssim 0.7$)
and for properties probing the velocity domain beyond the thermal
region, does the IHS system imprint its signature and distinguish
from the ``disguised'' EHS system. In this paper we have restricted
ourselves to the USF, but we plan to perform a similar comparison in
other states, especially in those where the heat flux, rather than
the pressure tensor, is the relevant quantity. We will also
undertake a parallel study in the case of dilute mixtures, as well as in the case
of dense gases (complemented by molecular dynamics simulations), following the schemes discussed in the preceding paper
\cite{SA04}.

\begin{acknowledgments}
We are grateful to V. Garz\'o  for a critical
reading of the manuscript.
Partial support from the Ministerio de
Educaci\'on y Ciencia (Spain) through grant No.\ FIS2004-01399
(partially financed by FEDER funds) is gratefully acknowledged.
A.A. is grateful to the Fundaci\'on Ram\'on Areces (Spain) for a predoctoral fellowship.
\end{acknowledgments}


\begin{thebibliography}{99}

\bibitem{GS95}  A. Goldshtein and M. Shapiro, {J. Fluid Mech.} {\bf 282},
75 (1995).

\bibitem{BDS97}
J. J. Brey, J. W. Dufty, and A. Santos, J. Stat. Phys. \textbf{87},
1051 (1997).

\bibitem{vNE01}
T. P. C. van Noije and M. H. Ernst, in \textit{Granular Gases}, edited by T.
P\"oschel and S. Luding,  Lecture Notes in Physics, Vol.\ 564
(Springer--Verlag, Berlin, 2001), pp.\ 3--30.

\bibitem{SA04}
A. Santos and A. Astillero, ``System of elastic hard
spheres which mimics the transport properties of a granular gas;'' arXiv: cond-mat/0405252.

\bibitem{G03}
I. Goldhirsch, Annu. Rev. Fluid Mech. \textbf{35}, 267 (2003).

\bibitem{SGD03}
A. Santos, V. Garz\'o, and J. W. Dufty, Phys. Rev. E \textbf{69},
061303 (2004).

\bibitem{BDS99}
J. J. Brey, J. W. Dufty, and A. Santos, J. Stat. Phys. \textbf{97},
281 (1999). Note a misprint in Eq.\ (C.11): The right-hand sides of
the two equations should be multiplied by $(d-1)(d+2)/2$ and $d$,
respectively.

\bibitem{BDKS98}
J. J. Brey, J. W. Dufty, C. S. Kim, and A. Santos, Phys. Rev. E
\textbf{58}, 4638 (1998).

\bibitem{CC70}
S. Chapman and T. G. Cowling, \textit{
 The Mathematical Theory of Nonuniform Gases} (Cambridge
University Press, Cambridge, 1970).

\bibitem{note1}
Note that, strictly speaking, the relationships $\lambda'={\lambda}/\beta$ and
$\tau'={\tau}/{\beta}$ only hold if the EHS and IHS  gases have the
same density and temperature.

\bibitem{note0}
This must be
distinguished from a system of \textit{rough} spheres with a
coefficient of normal restitution $\alpha=1$. The latter system has
been considered, for instance, by  S. J. Moon, J. B. Swift, and H. L.
Swinney, Phys. Rev. E \textbf{69}, 011301 (2004).

\bibitem{AS04}
For a preliminary report of this research, see A. Astillero and A.
Santos, ``A granular fluid modeled as a driven system of elastic
hard spheres,'' in \textit{The Physics of Complex Systems (New Advances and Perspectives)},
edited by F. Mallamace and H. E. Stanley (IOS Press, Amsterdam, 2004), pp.\ 475--480; arXiv: cond-mat/0309220.

\bibitem{LSJC84}
 {C. K. K. Lun, S. B. Savage, D. J.
Jeffrey, and N. Chepurniy}, {J. Fluid Mech.} {\bf 140}, 223 (1984).

\bibitem{CG86}
C. S. Campbell and A. Gong, J. Fluid Mech. {\bf 164}, 107 (1986).

\bibitem{WB86}
O. R. Walton and R. L. Braun, J. Rheol. \textbf{30}, 949 (1986);
Acta Mech. \textbf{63}, 73 (1986).

\bibitem{JR88}
{J. T. Jenkins and M. W. Richman} {J. Fluid Mech.} {\bf 192}, 313
(1988).

\bibitem{C89}
{C. S. Campbell}, {J. Fluid Mech.} {\bf 203}, 449 (1989).

\bibitem{C90}
C. S. Campbell, {Annu. Rev. Fluid Mech.} \textbf{22}, 57 (1990).

\bibitem{S92}
{S. B. Savage}, { J. Fluid Mech.} {\bf 241}, 109 (1992).

\bibitem{HS92}
{M. A. Hopkins and H. H. Shen}, { J. Fluid Mech.} {\bf 244}, 477
(1992).

\bibitem{SK94}
P. J. Schmid and H. K. Kyt\"omaa, J. Fluid Mech. \textbf{264}, 255
(1994).

\bibitem{LB94}
{C. K. K. Lun and A. A. Bent}, { J. Fluid Mech.} {\bf 258}, 335
(1994).

\bibitem{GT96}
{I. Goldhirsch and M. L. Tan}, {Phys. Fluids} {\bf 8}, 1752 (1996).

\bibitem{SGN96}
 {N. Sela, I.
Goldhirsch, and S. H. Noskowicz}, { Phys. Fluids} {\bf 8}, 2337
(1996).

\bibitem{C97}
C. S. Campbell, J. Fluid Mech. \textbf{348}, 85 (1997).

\bibitem{BRM97} J. J. Brey, M. J. Ruiz-Montero, and F. Moreno,
Phys. Rev. E \textbf{55}, 2846 (1997).

\bibitem{CR98}
C.-S. Chou and M. W. Richman, Physica A \textbf{259}, 430 (1998);
C.-S. Chou, \textit{ibid.} \textbf{287}, 127 (2000); \textbf{290},
341 (2001).

\bibitem{MGSB99}
J. M. Montanero, V. Garz\'o, A. Santos, and J. J. Brey, J. Fluid
Mech. \textbf{389}, 391 (1999).

\bibitem{K00}
V. Kumaran, Physica A \textbf{275}, 483 (2000); \textbf{284}, 246
(2000); \textbf{293}, 385 (2001).

\bibitem{F00}
A. Frezzotti, Physica A \textbf{278}, 161 (2000).

\bibitem{C01}
C. Cercignani, J. Stat. Phys. \textbf{102}, 1407 (2001).

\bibitem{AH01}
M. Alam and C. M. Hrenya, Phys. Rev. E \textbf{63}, 061308 (2001).

\bibitem{CH02}
R. Clelland and C. M. Hrenya, Phys. Rev. E \textbf{65}, 031301
(2002).

\bibitem{G02}
V. Garz\'o, Phys. Rev. E \textbf{66}, 021308 (2002); J. Stat. Phys.
\textbf{112}, 657 (2003).

\bibitem{MG02}
J. M. Montanero and  V. Garz\'o, Physica A \textbf{310}, 17 (2002);
Mol. Sim. \textbf{29}, 357 (2003).

\bibitem{BGS02}
A. V. Bobylev, M. Groppi, and G. Spiga, Eur. J. Mech. B/Fluids
\textbf{21}, 91 (2002).

\bibitem{GM03}
V. Garz\'o and J. M. Montanero, Phys. Rev. E \textbf{68}, 041302
(2003).

\bibitem{AL03}
M. Alam and S. Luding, J. Fluid Mech. \textbf{476}, 69 (2003);
Phys. Fluids \textbf{15}, 2298 (2003).


\bibitem{L04}
J. F. Lutsko, Phys. Rev. E \textbf{70}, 061101 (2004).

\bibitem{MGAS04}
J. M. Montanero, V. Garz\'o, M. Alam, and S. Luding,
``Rheology of granular mixtures under uniform shear flow: Enskog kinetic theory versus molecular dynamics
simulations,'' arXiv: cond-mat/0411548.

\bibitem{GS03}
See, for instance, V. Garz\'o and A. Santos, \textit{Kinetic Theory of Gases in Shear
Flows. Nonlinear Transport} (Kluwer Academic Publishers, Dordrecht,
2003), Chaps.\ 2--4, and references therein.

\bibitem{TTMGSD01} M. Tij, E. Tahiri, J. M. Montanero, V. Garz\'{o}, A. Santos,
and J. W. Dufty, J. Stat. Phys. \textbf{103}, 1035 (2001).

\bibitem{LE72}
A. W. Lees and S. F. Edwards, J. Phys. C \textbf{5}, 1921 (1972).

\bibitem{DSBR86}
J. W. Dufty, A. Santos, J. J. Brey, and R. F. Rodr\'{\i}guez, Phys.
Rev. A \textbf{33}, 459 (1986).

\bibitem{note2}
To avoid confusion in the notation, we use the superscript $0$ to
refer to initial quantities (e.g., $T^0$), while we usually employ the subscript
$0$ to denote local equilibrium quantities (e.g., $\zeta_0$).

\bibitem{B94}
G. Bird, \textit{Molecular Gas  Dynamics and the Direct Simulation
of Gas Flows} (Clarendon, Oxford, 1994).

\bibitem{AG97}
F. J. Alexander and A. L. Garcia, Comp. Phys. \textbf{11}, 588
(1997).

\bibitem{MS96}
For an extension of the DSMC method to the Enskog equation, see J.
M. Montanero and A. Santos,  Phys. Rev. E \textbf{54}, 438 (1996);
 Phys. Fluids \textbf{9}, 2057 (1997);
 A. Frezzotti, \textit{ibid.} \textbf{9}, 1329 (1997).

\bibitem{G05}
V. Garz\'o, ``Transport coefficients for a granular gas around
uniform shear flow,'' arXiv: cond-mat/0506013.

\bibitem{BGK54}
P. L. Bhatnagar, E. P. Gross, and M. Krook, Phys. Rev. \textbf{94}, 511
(1954);
P. Welander, Arkiv. Fysik \textbf{7}, 507 (1954).

\bibitem{vNE98}
T. P. C. van Noije and M. H. Ernst, {Gran. Matt.} \textbf{1}, 57 (1998).

\bibitem{MS00}
J. M. Montanero and A. Santos, Gran. Matt. \textbf{2}, 53 (2000).

\bibitem{BGP04}
A. V. Bobylev, I. M. Gamba, and V. A. Panferov, J. Stat. Phys. \textbf{116}, 1651 (2004).






\end{thebibliography}
\end{document}